\documentclass[twocolumn, apl, amsmath, amssymb, showpacs]{revtex4-2}
\usepackage{epsf}      
\usepackage{graphicx}
\usepackage{color}
\usepackage{soul}
\usepackage{gensymb}
\usepackage{sidecap}
\usepackage{amsmath}
\usepackage{mathtools}
\usepackage[colorlinks=true, linkcolor=blue, citecolor=blue, urlcolor=blue]{hyperref}

\begin{document}

\title{From Complex Magnetic Ground States to Magnetocaloric Effects: A Review of Rare Earth R$_2$In Intermetallic Compounds}
\author{Anis Biswas}
\email{anis@ameslab.gov}
\affiliation{Ames National Laboratory, U.S. Department of Energy, Iowa State University, Ames, Iowa 50011, USA} 
\author{Ajay Kumar} 
\email{ajay1@ameslab.gov }
\affiliation{Ames National Laboratory, U.S. Department of Energy, Iowa State University, Ames, Iowa 50011, USA} 
\author{Prashant Singh} 
\email{psingh84@ameslab.gov}
\affiliation{Ames National Laboratory, U.S. Department of Energy, Iowa State University, Ames, Iowa 50011, USA} 
 \author{Yaroslav Mudryk}
\email{slavkomk@ameslab.gov}
\affiliation{Ames National Laboratory, U.S. Department of Energy, Iowa State University, Ames, Iowa 50011, USA} 

\date{\today}

\begin{abstract}
R$_{2}$In (R = rare earth) intermetallics exhibit unusual magnetic and magnetocaloric properties, driven by subtle electronic effects, lattice distortions, and spin–lattice coupling. Most of these binary compounds adopt the hexagonal Ni$_{2}$In-type structure at room temperature, with Eu$_{2}$In and Yb$_{2}$In stabilizing in the orthorhombic Co$_{2}$Si-type lattice. Lighter lanthanide compounds Eu$_{2}$In, Nd$_{2}$In, and Pr$_{2}$In undergo first-order magnetic transitions with negligible hysteresis and minimal lattice volume change and exhibit giant cryogenic magnetocaloric effects, while heavy lanthanide R$_{2}$In compounds including Gd$_{2}$In show second-order transitions with moderate  magnetocaloric effect. No lanthanide-based  R$_{2}$In compound exhibits symmetry-breaking structural transition, while Y$_{2}$In transforms from hexagonal to orthorhombic structure near 250~K. Secondary low-temperature transitions, including spin reorientation or antiferromagnetic ordering, further enrich the magnetic phase landscape in these compounds. Integrating theoretical descriptors such as charge-induced strain and electronic structure provides predictive insight into phase stability and magnetocaloric performance, guiding the design of rare-earth intermetallics with tunable magnetic properties for cryogenic applications.\\

\end{abstract}

\maketitle

\section{ ~Introduction}

Rare-earth-based intermetallic compounds (RBIs) represent a fascinating class of magnetic materials that continue to attract significant research interest due to their complex and diverse chemistry and physics~\cite{Szytula1993, Ghazanfar2025, Petit2020, Kurumaji2019, Pecharsky1997, Chinchure2002, Asti1994, DelRose2021, Ribeiro2013, Guillou2018, Biswas2020, Ahmed2025, Bhattacharya2025, Biswas2024, Alho2022, Chakraborty2022, Biswas2024b, Kumar2024, Chakraborty2024, Ribeiro2024, Kumar2025a, Kumarprb24, Mudryk2011, Custers2003, Paschen2004, Custers2012, EuAl4Skyrmions2022, Caretta2020, Zhang2021, Kumar2024b, Schuberth2016, Kenzelmann2008, Dong2025, Izawa2007, Bouaziz2022, Gd2PdSi3PRL2025}. These materials often exhibit fundamentally interesting and technologically important magnetic behaviors owing to their strong electron correlations, high magnetic moments, considerable spin-orbit coupling, and a common interplay between magnetic and structural degrees of freedom, features that remain central to contemporary condensed matter research~\cite{Szytula1993, Ghazanfar2025, Petit2020, Kurumaji2019, Pecharsky1997, Chinchure2002, Asti1994, DelRose2021, Ribeiro2013, Guillou2018, Biswas2020, Ahmed2025, Bhattacharya2025, Biswas2024, Alho2022, Chakraborty2022, Biswas2024b, Kumar2024, Chakraborty2024, Ribeiro2024, Mudryk2011, Kumar2025a, Kumarprb24, Custers2003, Paschen2004, Custers2012, EuAl4Skyrmions2022, Caretta2020, Zhang2021, Kumar2024b, Schuberth2016, Kenzelmann2008, Dong2025, Izawa2007, Bouaziz2022, Gd2PdSi3PRL2025}. These distinctive characteristics of RBIs make them ideal materials for both current, primarily as permanent magnets, and future applications ranging from high-density data storage and spintronics to magnetic refrigeration and renewable energy technologies~\cite{Ghazanfar2025,  Pecharsky1997, Chinchure2002, Asti1994, DelRose2021, Guillou2018, Biswas2020, Zhang2021, Kumar2024b, Izawa2007, Nakamurascripta18}.

The complexity of these materials arises largely from the presence of 4$f$ electrons, common for most rare earths (except Sc, Y, and La), which adds intricacy to the interplay between chemical composition, crystal structure, and physical behavior. The 4$f$ electrons are typically localized, leading to magnetic interactions that are predominantly mediated by conduction electrons via the indirect exchange mechanism, commonly known as Ruderman-Kittel-Kasuya-Yosida (RKKY) model~\cite{Jensen1991, Kasuya1956,Campbell1972}.  The RKKY model, however, is too simple to quantitatively estimate complex interplay between multiple magnetic exchange interactions in complex RBIs with a variety of interatomic distances. Moreover, hybridization between the localized 4$f$ states and the more extended 5$d$ orbitals adds an additional layer of complexity in describing magnetic interactions in these compounds. On top of this already complicated scenario, the role of additional $s$-, $p$-, and $d$-electrons brought by alloying with non-rare-earth atoms cannot be overstated, even when the added constituents are non-magnetic. These delocalized electrons that also participate in chemical bonding can strongly influence the effective exchange coupling of magnetic atoms, introduce magnetocrystalline anisotropy, and crystalline electric field splitting of 4$f$ energy levels by virtue of modifying atomic environments. In turn, alloying can also alter the nature of magnetic phase transitions~\cite{Biswas2024b, Mudryk2011, Kumar_PRB_25}. Investigation of such systems often uncovers new physical phenomena, which, consequently, may lead to novel magnetic functionalities.

 \begin{figure*}
\includegraphics[width=7in]{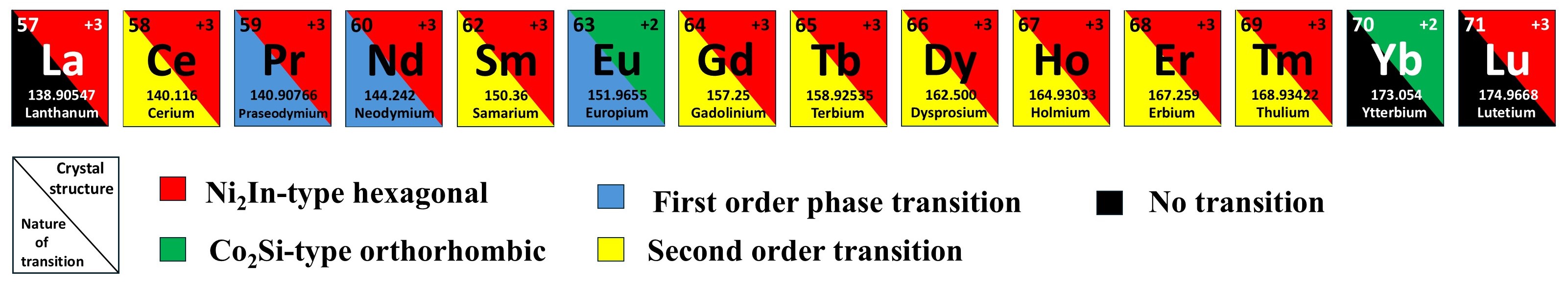} 
\caption {Crystal structure (top-right half, color-coded) and nature of the magnetic phase transition (bottom-left half, color-coded) of R$_2$In compounds for different rare-earth elements. The numbers in the top-right corners indicate the valence state of the rare-earth ions in the corresponding compounds.}
\label{Fig1}
\end{figure*}

Among the broad class of rare earth-based intermetallic compounds, those with the general formula R$_2$In (where R is a rare earth element) have drawn considerable attention for their recently uncovered intriguing structural, magnetic, and electronic behaviors, influencing fundamental condensed matter research and showing potential for technological applications~\cite{Guillou2018, Biswas2020,  Forker2005, Bhattacharyya2009, Zhang2009, Ravot1993, Zhang2011, Alho2020, Ryan2019, MendiveTapia2020, Singh2012, Bhattacharyya2012, Tence2016, McAlister1984a, Kamali2014, Taskaev2021, Liu1987, Jee1996, Bhattacharyya2009b, Zhang2009b, Ravot1992, Ravot1997, Biswas2022a, Biswas2022b, Liu2021, Cui2022, GamariSeale1979, Parviainen1980, Bazela1988, Ravot1995, McAlister1984b, OlzonDionysio1992, Zhang2009c, Liu2024, Svanidze2008, Guillou2020, Biswas2023, GegenMater25}. These compounds typically crystallize in either orthorhombic (for R = Eu and Yb) or hexagonal structures for the rest of the rare-earth elements (see Fig.~\ref{Fig1})~\cite{Guillou2018, Biswas2020, Bazela1988}. A key feature of many R$_2$In compounds is the onset of magnetic phase transition at cryogenic temperatures, often accompanied by a pronounced magnetocaloric effect (MCE)~\cite{Guillou2018, Biswas2020,  Bhattacharyya2009, Zhang2009, Zhang2011, Tence2016, Bhattacharyya2009b, Zhang2009b, Biswas2022b, Liu2021, Cui2022, Zhang2009c, Liu2024, Biswas2023, GegenMater25}.  Notably, recent studies have revealed an especially interesting phenomenon—the presence of first-order magnetic phase transition (FOMT) with negligibly small thermal hysteresis in members with R = Pr, Nd, and Eu (see Fig.~\ref{Fig1}), accompanied by invariance of the crystal symmetry and a very small ($\leqslant$ 0.1\%) change in the unit cell volume at the FOMT~\cite{Guillou2018, Biswas2020, Biswas2022b, Liu2021}. Such transitions, which are uncommon yet highly desired due to reversibility and lack of energy losses, elevate the scientific interest in R$_2$In compounds. In addition to the well-established usefulness of such transitions for next-generation magnetic refrigeration, a fundamental understanding of the mechanisms responsible for negligible thermal hysteresis associated with them will impact other advanced technological applications that utilize magnetic phase changes.

This review provides a comprehensive overview of the \(R_2\)In family of rare-earth intermetallic compounds, with a particular focus on their crystallographic, magnetic, and magnetocaloric properties. We begin by discussing the structural variations across this rare-earth series, followed by describing intrinsic magnetic behavior and the nature of magnetic phase transitions exhibited by these materials. A dedicated section explores the magnetocaloric effect (MCE), emphasizing compounds that demonstrate substantial isothermal entropy changes near the boiling points of technologically useful gases, such as hydrogen ($\sim$20~K), oxygen ($\sim$90~K)  and natural gas ($\sim$110~K). These observations highlight the potential of \(R_2\)In compounds for use in energy-efficient cooling systems and gas liquefaction technologies. In addition to the key experimental findings, the review integrates insights from theoretical modeling and computational studies that shed light on the complex interplay among electronic structure, crystallography, and magnetic interactions. These perspectives are critical for the rational design and optimization of materials with improved magnetocaloric and other functional aspects of the materials.

Finally, we outline several promising avenues for future research. One direction involves chemical substitution strategies, aimed at systematically tuning the magnetic interactions and magnetocaloric responses of \(R_2\)In compounds through controlled modification of rare-earth or transition-metal sites. Complementary to this, structural engineering approaches such as strain induction, dimensional confinement, or controlled defect incorporation, may provide additional pathways to enhance functional properties, including magnetocrystalline anisotropy, Curie temperature (\(T_C\)), and thermal stability. Despite substantial progress, key scientific questions remain unresolved, including the detailed mechanisms underlying first-order magnetic transitions, the role of electronic correlations in magnetocaloric behavior, and the interplay between lattice distortions and magnetic ordering. Addressing these challenges will be critical to fully understanding the rich physics of the \(R_2\)In family and leveraging their potential in high-performance magnetic and energy-conversion applications.

 \begin{figure*}
\includegraphics[width=7in]{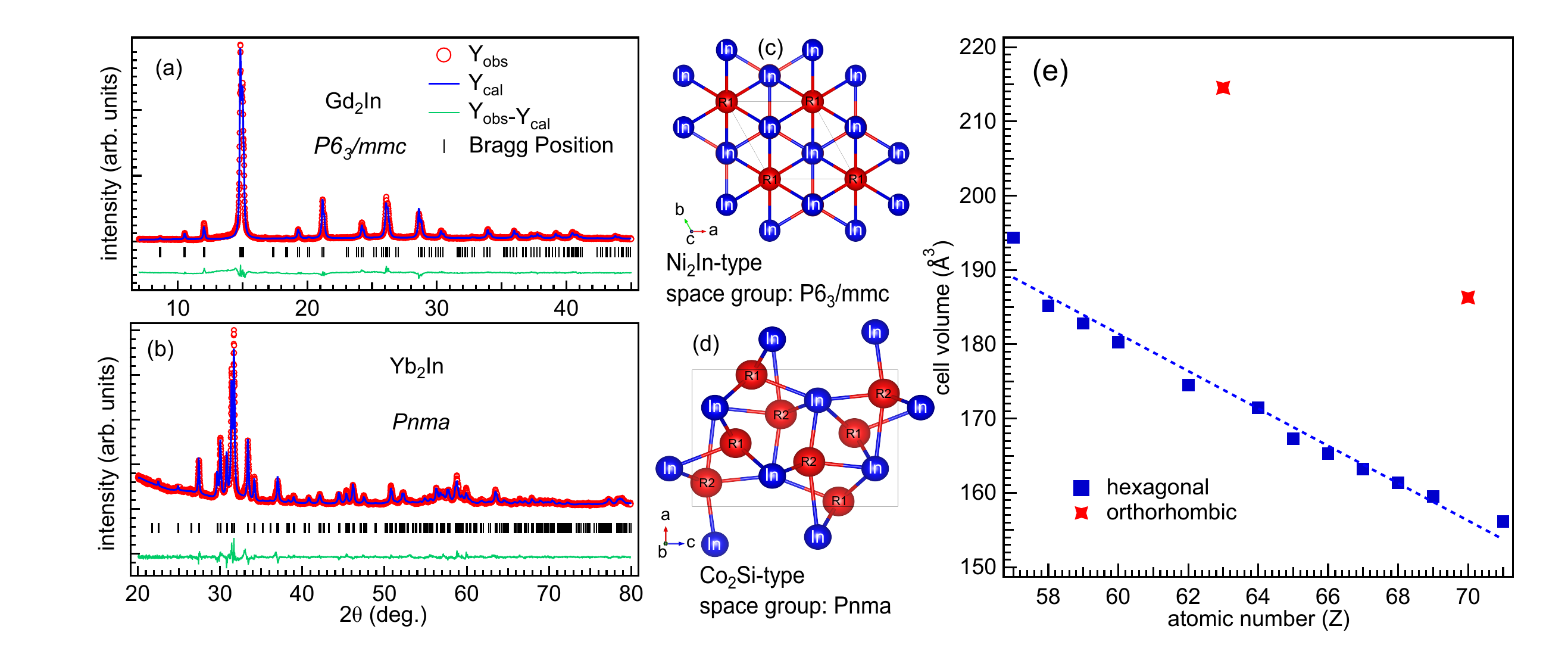} 
\caption {(a) and (b) Room-temperature X-ray diffraction patterns for Gd$_2$In and Yb$_2$In, collected  using Mo K$_\alpha$ and Cu K$_\alpha$ radiation, respectively: representative of Ni$_2$In-type hexagonal and Co$_2$Si-type orthorhombic structures adopted by R$_2$In. Figures (a, b) are reprinted with permission from \cite{Singh2025}, Copyright (2025) by the American Physical Society. (c, d) Corresponding Ni$_2$In hexagonal and Co$_2$Si orthorhombic unit cells (red = R, blue = In), with rare-earth atoms on two inequivalent sites. (e) Experimental unit-cell volume versus atomic number $Z$ for the lanthanide series: unit-cell volume of hexagonal compounds (blue) decreases monotonically from La to Lu, while the volume of orthorhombic Eu$_2$In and Yb$_2$In cells (red; normalized cell volume shown, i.e., half the full cell) is much higher but follows the same contracting trend. The lattice volume values are taken from Table~\ref{tab:structure}.  }
\label{Fig2}
\end{figure*}

\section{\noindent ~Discussion }

\subsection{\noindent ~Sample preparation and handling }

All hexagonal R$_2$In compounds reported in the literature were synthesized using the conventional arc-melting technique under an argon atmosphere \cite{Zhang2011, Bhattacharyya2009b, Bhattacharyya2009, Zhang2009c, Singh2012, Biswas2020, Biswas2022b, Biswas2023, Liu2021}. Stoichiometric amounts of the constituent elements were arc-melted  several times, with the sample button flipped between melts to ensure homogeneity. The as-prepared samples were subsequently annealed for several days in evacuated or He-filled sealed quartz tubes at temperatures between 700 and 800$^\circ$C, depending on the composition \cite{Zhang2011, Bhattacharyya2009b, Bhattacharyya2009, Zhang2009c, Biswas2020, Biswas2022b, Biswas2023}. For compounds based on heavy lanthanides, the annealing duration typically ranged from a few days up to, but not exceeding, one week \cite{Zhang2011, Bhattacharyya2009b, Bhattacharyya2009, Zhang2009c, Singh2012}. In contrast, for certain light lanthanide-based compounds, such as Pr$_2$In and Nd$_2$In, significantly longer annealing times---up to three weeks---have been reported \cite{Biswas2020, Biswas2022b, Biswas2023}. In comparison, orthorhombic R$_2$In compounds (R = Eu and Yb) were synthesized by melting the constituent elements in sealed tantalum crucibles under a partial atmosphere of ultrapure argon using a resistive furnace between 900 and 1000$^\circ$C \cite{Guillou2018, Guillou2020}. This approach was employed to minimize rare-earth evaporation during the melting process. During synthesis, the crucible was enclosed within a highly pure helium-filled quartz tube to prevent oxidation of the tantalum at elevated temperatures \cite{Guillou2018, Guillou2020}. Following melting, these compounds required annealing at lower temperatures, typically between 650 and 700 $^\circ$C, for 1–2 days \cite{Guillou2018, Guillou2020}.

Most light rare-earth-based R$_2$In compounds, as well as Yb$_2$In and the non-lanthanide compound Y$_2$In, are highly air-sensitive and unstable under ambient conditions \cite{Biswas2020, Biswas2022b, Biswas2023, Guillou2018, Guillou2020, Ravot1995, Svanidze2008, Forker2005}. Consequently, all sample handling was carried out in an argon-filled glovebox. In this context, Forker \textit{et al.}\ reported significant difficulties in obtaining reliable X-ray diffraction (XRD) data for Pr$_2$In due to its rapid oxidation, even when powders were prepared in a glovebox and protected with silicone grease \cite{Forker2005}. By contrast, successful XRD measurements on several light rare-earth-based R$_2$In compounds, including Pr$_2$In, have been achieved by preparing and mounting the powders entirely inside a glovebox and subsequently covering them with a thin Kapton film \cite{Biswas2020, Biswas2022b, Biswas2023} or sealed within a Kapton capillary \cite{Biswas2022a}. It is worth noting that  Nd$_2$In exhibits better air stability compared to Pr$_2$In and hence partial substitution of Pr by Nd improves the ambient stability of Pr$_2$In  \cite{Biswas2023}.

\subsection{\noindent ~Crystal structure of R$_2$In }

R$_2$In compounds are reported to form with all rare earth elements, except unstable Pm, and their lattice parameters and structure types are reported in Table \ref{tab:structure}.  Among all R$_2$In compounds where R is a lanthanide, only two---Eu$_2$In and Yb$_2$In---crystallize in the Co$_2$Si orthorhombic structure (space group \textit{Pnma}), while the remaining members of the family adopt a Ni$_2$In-type hexagonal structure (space group \textit{P6$_3$/mmc}) at room temperature~\cite{Guillou2018, Biswas2020, Bazela1988, Singh2025}. The lattice volume of the hexagonal structure systematically decreases from light to heavy lanthanide elements [Fig.~\ref{Fig2}(e) and Table~\ref{tab:structure}] in accordance with the lanthanide contraction~\cite{Bazela1988, Singh2025}. A significantly larger normalized volume per formula unit is observed for the orthorhombic structure as compared to the hexagonal one [Fig.~\ref{Fig2}(e)]. Notably, none of the R$_2$In compounds with R belonging to the lanthanide series undergo a change in crystal symmetry upon cooling at the magnetic ordering temperature; they retain their room-temperature structure as the ground state, even though they may exhibit variations in lattice parameters and unit cell volume near their magnetic transition temperatures~\cite{Guillou2018, Biswas2020, Biswas2022a, Biswas2022b, Singh2025}. A notable exception is Y$_2$In, a R$_2$In compound based on the non-lanthanide rare earth (Y), which transforms from the hexagonal (room temperature) to the orthorhombic (low temperature) structure around 250~K~\cite{Svanidze2008}. It is worth mentioning that in both hexagonal and orthorhombic structures, rare-earth atoms reside in two symmetry-distinctive sites as shown in Figs.~\ref{Fig2}(a--d), respectively~\cite{Guillou2018, Biswas2020, Biswas2022b}.

\begin{table*}
\centering
\caption{Crystal structure types and lattice parameters of $R_2$In binary compounds at room temperature$^@$.}
\label{tab:structure}
\begin{tabular}{lllllll}
\hline\hline
\textbf{Compound} & \textbf{Crystal structure} & $a$ (\AA) & $b$ (\AA) & $c$ (\AA) & $V$ (\AA$^3$) & \textbf{Reference} \\ \hline

Sc$_2$In & Hex-Ni$_2$In & 5.05 & 5.05 & 6.30 & 139.14 & \cite{Wiendlocha2006} \\ \hline

Y$_2$In  & Hex-Ni$_2$In & 5.3599(2) & 5.3599(2) & 6.7647(3) & 168.57 & \cite{Svanidze2008} \\
         & Ortho-Co$_2$Si (at 200 K)$^{\rm @}$ & 6.7486(4) & 5.1426(4) & 9.7382(7) & 337.97 & \cite{Svanidze2008}\\ \hline

La$_2$In & Hex-Ni$_2$In & 5.636 & 5.636 & 7.065 & 194.3501 & \cite{Palenzona1968} \\\hline
Ce$_2$In & Hex-Ni$_2$In & 5.562 & 5.562 & 6.911 & 185.1542 & \cite{Palenzona1968} \\
         &               & 5.47(1) & 5.47(1) & 6.820(1) & 176.7(1) & \cite{Bazela1988} \\\hline
Pr$_2$In & Hex-Ni$_2$In & 5.534 & 5.534 & 6.893 & 182.8173 & \cite{Palenzona1968} \\
         &               & 5.528(6) & 5.528(6) & 6.878 & 182.024 & \cite{Cui2022} \\
         &               & 5.539(6), 5.539(2) & 5.539(6), 5.539(2) & 6.902(10), 6.895(3) & 183.254, 183.201 & \cite{Biswas2020}$^*$\\
         &               & 5.484(1) & 5.484(1) & 6.840(2) & 178.148 & \cite{Liu2024} \\
         &               & 5.5335 & 5.5335 & 6.8969 & 182.887 & \cite{Biswas2022a} \\\hline
Nd$_2$In & Hex-Ni$_2$In & 5.505 & 5.505 & 6.868 & 180.2501 &  \cite{Palenzona1968} \\
         &               & 5.490(1) & 5.490(1) & 6.850(1) & 178.80 & \cite{Bazela1988} \\
         &               & 5.5085(7) & 5.5085(7) & 6.8745(10) & 180.650 & \cite{Biswas2022b} \\
         &               & 5.491(3) & 5.491(3) & 6.856(4) & 178.989 & \cite{Cui2022} \\\hline
Sm$_2$In & Hex-Ni$_2$In & 5.450 & 5.450 & 6.785 & 174.5314 & \cite{Palenzona1968}\\\hline
Eu$_2$In & Ortho-Co$_2$Si & 7.4536(6) & 5.5822(5) & 10.312(1) & 429.056 & \cite{Guillou2018}\\
         &               & 7.445(2) & 5.573(1) & 10.306(4) & 427.61 & \cite{Fornasini1990} \\\hline
Gd$_2$In & Hex-Ni$_2$In & 5.413 & 5.413 & 6.756 & 171.4338 & \cite{Palenzona1968}  \\
         &               & 5.410(1) & 5.410(1) & 6.751(1) & 170.7(1) & \cite{Bazela1988} \\
         &               & 5.413 & 5.413 & 6.750 & 178.282 & \cite{Singh2012}\\
         &               & 5.413 & 5.413 & 6.756 & 171.4 & \cite{Tence2016} \\
         &               & 5.461 & 5.461 & 6.771 & 174.875 & \cite{Taskaev2021} \\
         &               & 5.41 & 5.41 & 6.76 & 171.345 & \cite{Jee1996} \\\hline
Tb$_2$In & Hex-Ni$_2$In & 5.367 & 5.367 & 6.707 & 167.3101 & \cite{Palenzona1968} \\
         &               & 5.360(7) & 5.360(7) & 6.702(5) & 166.7(6) & \cite{Bazela1988} \\
         &               & 5.367 & 5.367 & 6.707 & 167.310 & \cite{OlzonDionysio1992} \\ \hline

Dy$_2$In & Hex-Ni$_2$In & 5.346 & 5.346 & 6.677 & 165.2608 & \cite{Palenzona1968}\\
         &               & 5.340 & 5.340 & 6.660 & 164.470 & \cite{Bhattacharyya2009} \\
         &               & 5.349(1) & 5.349(1) & 6.675(1) & 165.4(1) &\cite{Bazela1988} \\ \hline

Ho$_2$In & Hex-Ni$_2$In & 5.319 & 5.319 & 6.662 & 163.2282 & \cite{Palenzona1968} \\
         &               & 5.314(3) & 5.314(3) & 6.670(5) & 163.1(3) & \cite{Bazela1988}  \\
         &               & 5.300 & 5.300 & 6.660 & 162.0155 & \cite{Bhattacharyya2009b} \\
         &               & 5.3216(10) & 5.3216(10) & 6.6667(9) & 163.5031 & \cite{Anh2006} \\
         &               & 5.301 & 5.301 & 6.624 & 161.201 & \cite{Ravot1992} \\ \hline

Er$_2$In & Hex-Ni$_2$In & 5.297 & 5.297 & 6.641 & 161.3705 & \cite{Palenzona1968}\\
         &     & 5.295(3) & 5.295(3) & 6.580(5) & 161.4(3) &  \cite{Bazela1988} \\
         &     & 5.2909(6) & 5.2909(6) & 6.6373(9) & 160.9093 & \cite{Ravot1993} \\ \hline

Tm$_2$In & Hex-Ni$_2$In & 5.274 & 5.274 & 6.621 & 159.4904 & \cite{Palenzona1968}\\ \hline

Yb$_2$In & Ortho-Co$_2$Si & 7.093(1) & 5.3296(7) & 9.884(2) & 373.643 & \cite{Guillou2020} \\
         &                 & 7.072(2) & 5.340(1) & 9.866(5) & 372.584 &  \cite{McMasters1971}\\ \hline

Lu$_2$In & Hex-Ni$_2$In & 5.239 & 5.239 & 6.569 & 156.1445 & \cite{Palenzona1968} \\ \hline\hline

\end{tabular}

\begin{flushleft}
\footnotesize{$^*$Lattice parameters of two different samples of Pr$_2$In prepared via different methods  as  reported in Ref. \cite{Biswas2020}.\\
$^{\rm @}$Crystal structure and lattice parameters of Y$_2$In at 200~K are also presented. }
\end{flushleft}
\end{table*}

 \begin{figure*}
\includegraphics[width=5in]{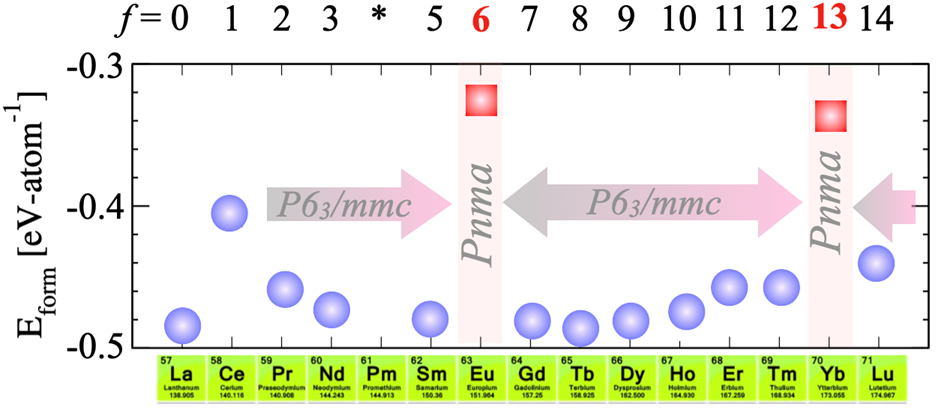} 
\caption {Formation energies ($E_{\mathrm{form}}$) per atom for R$_2$In across the lanthanide series (La $\rightarrow$ Lu). Blue circles mark calculated $E_{\mathrm{form}}$ (eV$\cdot$atom$^{-1}$) for compounds relaxed in the hexagonal ($P6_3/mmc$) motif, while the two red squares at $f = 6$ and $f = 13$ indicate much less negative enthalpies and a ground-state switch to an orthorhombic ($Pnma$) structure (shaded columns). The abrupt energy increase at these two $f$-counts ($\approx 0.15$–0.18 eV/atom relative to neighbors) signals $f$-electron/valence-driven changes in bonding (e.g., divalent character or $f$-localization) and coincides with the annotated $P6_3/mmc$ $\rightarrow$ $Pnma$ structural change. Figure is reprinted  with permission from \cite{Singh2025}, Copyright (2025) by the American Physical Society. }
\label{Fig3}
\end{figure*}

To clarify the ground-state stability of crystal structure in R$_2$In compounds, Figure~\ref{Fig3} reproduces the phase-stability results from Singh \textit{et al.} \cite{Singh2025}, showing Density functional theory (DFT) based formation energies, plotted as a function of the 4$f$-electron count. Notably, most R$_2$In compounds in the La $\rightarrow$ Lu row have negative formation energies around $-0.45$ to $-0.50$ eV/atom, indicating a clear stability, while two clear outliers appear near $f = 6$ and $f = 13$. Those two points (red squares) have much less negative formation energies ($-0.32$ to $-0.34$ eV/atom), a change on the order of $0.15$–$0.18$ eV/atom compared with their neighbors, and in those compositions the crystal structure changes from the hexagonal $P6_3/mmc$ to an orthorhombic $Pnma$. Because the anomalies align with the specific $f$-electron numbers, the most natural interpretation is that filling of the $f$-electron shell dictates changes in valence and localization of the $f$ orbitals, which strongly modifies bonding and therefore structural preference.  

Across the lanthanide series, the DFT-calculated formation energy shows clear anomaly for Eu$_2$In and Yb$_2$In compounds (Fig.~\ref{Fig3}). This behavior can be attributed to electron transfer from the 5$d^1$ state to the 4$f^6$/4$f^{13}$ states, resulting in the more stable 4$f^7$/4$f^{14}$ configurations. These changes in the 5$d$ state together with the significant increase in atomic size of now divalent rare-earth atoms promote stabilization of the orthorhombic $Pnma$ phase in Eu$_2$In and Yb$_2$In, respectively. This highlights the significant influence of 4$f$–5$d$ electron transfer on the structural phase stability of R$_2$In compounds \cite{Singh2025}. To further elaborate, such valency change (e.g., divalent behavior of Eu or Yb) or abrupt changes in the effective ionic radius due to lanthanide contraction can modify R–In bond lengths and packing and thus destabilize the hexagonal structure in favor of an orthorhombic distortion \cite{Singh2025}. Localized $f$-electrons and associated magnetic/orbital interactions also alter the total energy and may favor a lower-symmetry lattice that better accommodates the electronic or magnetic state. 

\subsection{\noindent ~  Magnetic properties and phase transitions of R$_2$In}

\begin{figure*}
\includegraphics[width=7in]{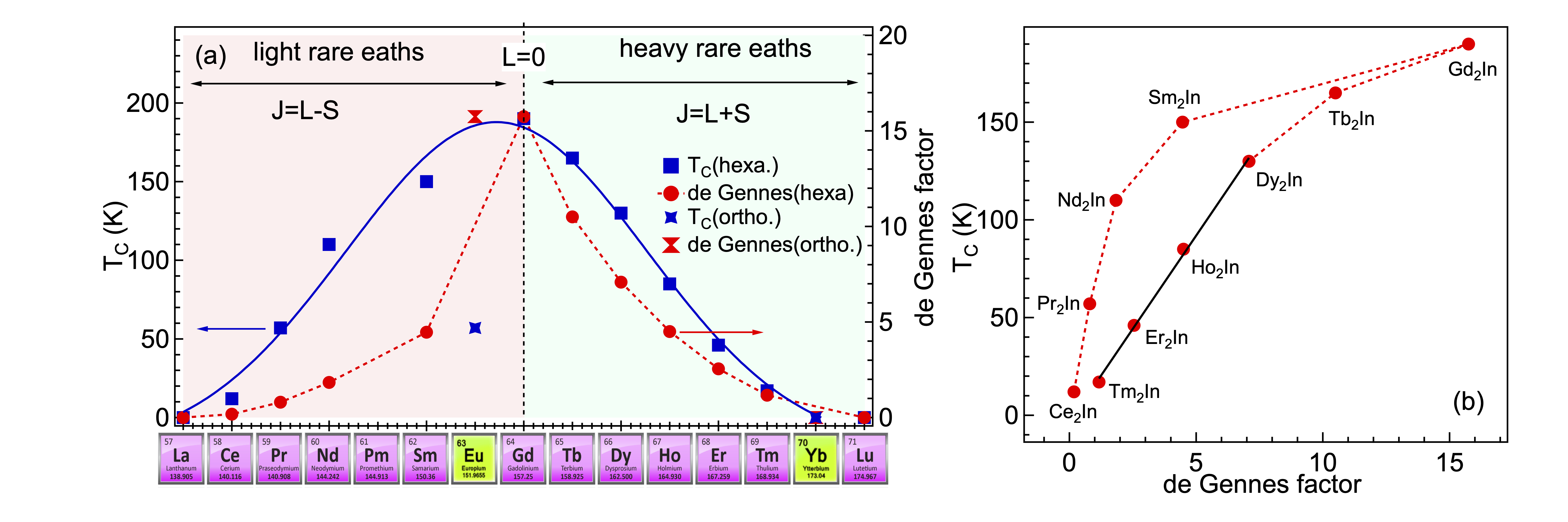} 
\caption {(a) Variation of the $T_{\mathrm{C}}$ of R$_2$In compounds (blue dots) and the de Gennes factor (dG) of the corresponding rare-earth elements (red dots) with atomic number ($Z$). For Eu$_2$In ($Z = 63$) and Yb$_2$In ($Z = 70$), dG is calculated assuming the divalent state of rare-earth elements, whereas for all other compounds, the tri-valency of rare-earths is considered. (b) $T_{\mathrm{C}}$ versus de Gennes factor for the hexagonal R$_2$In compounds. Here transition temperatures of R$_2$In compounds are taken from Refs. \cite{Guillou2018, Biswas2020, Forker2005, Bhattacharyya2009, Ravot1993, Singh2012, Bhattacharyya2009b, Biswas2022b, McAlister1984b, Chen1989}.}
\label{Fig4}
\end{figure*}

Depending on the choice of R, R$_2$In compounds exhibit various types of phase transitions leading to complex magnetic states. It is well established that the magnetic ordering temperatures ($T_{\mathrm{C}}$) of rare-earth intermetallics generally scale with the de Gennes factor, defined as dG $= (g - 1)^2 J(J + 1)$, where $g$ is the Landé $g$-factor and $J$ is the total angular momentum quantum number \cite{Singh2023}. In an earlier study, Forker \textit{et al.} reported an approximately linear relation between transition temperatures (extracted from the hyperfine field experiment) and dG for R$_2$In compounds in cases of both heavy as well as light rare-earth elements \cite{Forker2005}.  In Figure~\ref{Fig4}(a), we plot the $T_{\mathrm{C}}$ of R$_2$In compounds alongside the de Gennes factor of the corresponding rare-earth elements as a function of atomic number ($Z$). Both $T_{\mathrm{C}}$ and dG exhibit a maximum for Gd$_2$In and decrease away from it. The variation of $T_{\mathrm{C}}$ of different hexagonal R$_2$In compounds with dG of respective trivalent rare-earths [Fig.~\ref{Fig4}(b)] shows that $T_{\mathrm{C}}$ linearly scales with dG for only four R$_2$In compounds (R = Tm, Er, Ho, and Dy) where the transition is second-order ferromagnetic. However, a notable deviation from the linear scaling relation has been observed for compounds with $Z < 66$, possibly because of either the onset of competition between antiferromagnetic and ferromagnetic states at low temperature (for Tb$_2$In, Gd$_2$In, and Sm$_2$In) \cite{Zhang2009c, Bhattacharyya2012, McAlister1984b} or because the nature of the transition changes to first-order (Pr$_2$In, Nd$_2$In) \cite{Biswas2020, Biswas2022b, Liu2021}.  The reported ordering temperatures and other magnetic parameters of R$_2$In compounds are given in Table~\ref{tab:tc}.

To understand the role of rare-earth elements on magnetic properties and phase transitions in R$_2$In, Forker \textit{et al.} explored the hyperfine interactions—both magnetic and electric—of the probe isotope $^{111}$Cd situated on the indium site in a series of R$_2$In compounds (with R = Pr, Nd, Sm, Gd, Dy, Ho, Er, and Tm), using perturbed angular correlation (PAC) spectroscopy \cite{Forker2005}. According to their study, within the paramagnetic regime, the observed quadrupole interaction shows a substantial decline—over an order of magnitude—across the series from Pr to Er, highlighting the significant role of 4$f$ electron configurations in influencing the local charge environment at the indium site \cite{Forker2005}.

In the magnetically ordered phase, comprehensive measurements were carried out to evaluate the temperature and spin behavior of the magnetic hyperfine field, including its alignment with respect to the crystallographic $c$-axis of the hexagonal structure \cite{Forker2005}. Comparison between the hyperfine fields of $^{111}$Cd in the R$_2$In compounds and in elemental rare-earth metals suggests that the indirect coupling between 4$f$ electrons is primarily mediated through intra-atomic 4$f$–5$d$ exchange and interatomic 5$d$–5$d$ interactions, rather than through $s$-conduction electron mechanisms. Differences in the strength of the $f$–$d$ exchange interaction, characterized by the exchange parameter $\tau$, are evident between lighter (Pr, Nd, Sm) and heavier (Gd, Tb, etc.) rare-earth elements, with the former exhibiting significantly stronger coupling \cite{Forker2005}. 

Interestingly, among the systems studied, only Pr$_2$In and Nd$_2$In exhibit an abrupt discontinuity in the hyperfine field at the magnetic transition temperature—an indication of a first-order magnetic phase transition \cite{Forker2005}.  In contrast, the remaining compounds display a smooth, continuous change in the hyperfine field. Near the transition temperature, an observable broadening of the magnetic interaction linewidth points to spatial inhomogeneities in the magnetic exchange and a distributed range of local Curie temperatures. Additionally, changes in the angular orientation of the 4$f$ magnetic moments relative to the crystal’s $c$-axis as a function of temperature were inferred from variations in the hyperfine field direction with respect to the principal axis of the electric field gradient \cite{Forker2005}.

\begin{table*}
\centering
\caption{Magnetic and magnetocaloric parameters for R$_2$In compounds.}
\label{tab:tc}
\begin{tabular}{p{1.75cm}p{5cm}p{1.75cm}p{1.75cm}p{3cm}p{2cm}p{1.5cm}}
\hline\hline
Compound & Transition temperature & $\mu_{\mathrm{eff}}$ ($\mu_{\mathrm{B}}$/f.u.) & $M_{\mathrm{s}}$ ($\mu_{\mathrm{B}}$/f.u.) & -$\Delta S_{\rm Max}$(J/kg.K) \{$\Delta$H (kOe)\}& $\Delta T_{\rm ad}$(K) \space \{$\Delta$H (kOe)\} & References \\
\hline
Ce$_2$In  & $T_{\mathrm{N}} = 12$ K & 4.143 &  & & & \cite{Bazela1988,Chen1989} \\\hline

Pr$_2$In  & $T_{\mathrm{C}} = 57$ K, $T_{\mathrm{SR}} = 35$ K & 5.2  & 3.2 &15 \{20\} & 2 \{20\} & \cite{Biswas2020, Liu2024} \\
          & $T_{\mathrm{C}} = 57$ K, $T_{\mathrm{SR}} = 38$ K &      &   & &   & \cite{Cui2022}  \\\hline
Nd$_2$In  & $T_{\mathrm{C}} = 110$ K, $T_{\mathrm{SR}} = 53$ K & 5.18 & 4.94 & 13 \{20\} & & \cite{Biswas2022b} \\

 & $T_{\mathrm{C}} = 109$ K, $T_{\mathrm{SR}} = 50$ K &      &      & & 1.13 \{19.5\} & \cite{Liu2021}\\\hline
Sm$_2$In  & $T_{\mathrm{C}} = 152$ K & & & & & \cite{McAlister1984b} \\\hline
Eu$_2$In  & $T_{\mathrm{C}} = 55$ K & 12.02 & 14.4 &  26 \{20\} & 5 \{20\} & \cite{Guillou2018} \\\hline
Gd$_2$In  & $T_{\mathrm{C}} = 187$ K, $T_{\mathrm{N}} = 100$ K & & & & & \cite{Singh2012} \\
          & $T_{\mathrm{C}} = 190$ K, $T_{\mathrm{N}} = 105$ K & & & & & \cite{Bhattacharyya2012} \\
          & $T_{\mathrm{C}} = 187$ K, $T_{\mathrm{N}} = 99.5$ K & & & & & \cite{McAlister1984a} \\
          & $T_{\mathrm{C}} = 194$ K, $T_{\mathrm{N}} = 95$ K & & & 3.25 \{30\} & & \cite{Taskaev2021} \\
          & $T_{\mathrm{C}} = 190$ K, $T_{\mathrm{N}} = 100$ K & 10.06 & & & & \cite{Jee1996} \\\hline
Tb$_2$In  & $T_{\mathrm{C}} = 165$ K, $T_{\mathrm{N}} = 45$ K & & & 6.6 \{50\} & & \cite{Zhang2009c}\\\hline
Dy$_2$In  & $T_{\mathrm{C}} = 125$ K, $T_{\mathrm{SR}} = 50$ K & 15.6 & 17 & 7.1 \{40\} & & \cite{Bhattacharyya2009} \\
          & $T_{\mathrm{C}} = 130$ K & & & & & \cite{Zhang2009} \\\hline
Ho$_2$In  & $T_{\mathrm{C}} = 85$ K, $T_{\mathrm{SR}} = 32$ K & 15 & & & & \cite{Bhattacharyya2009b} \\
          & $T_{\mathrm{C}} = 85$ K, $T_{\mathrm{SR}} = 32$ K & & & 11.2 \{50\} & & \cite{Zhang2009b}\\\hline
Er$_2$In  & $T_{\mathrm{C}} = 40$ K & & 16 & & & \cite{Ravot1993} \\
          & $T_{\mathrm{C}} = 46$ K & 13.58 & & 16 \{50\} & & \cite{Zhang2011} \\\hline
Tm$_2$In  & $T_{\mathrm{C}} = 17$ K & 10.75 & & & & \cite{Forker2005,GamariSeale1979} \\\hline
Yb$_2$In  & Diamagnetic & & & & &\cite{Guillou2020} \\\hline
Y$_2$In  & Paramagnetic & & & & & \cite{Svanidze2008} \\\hline
\hline
\end{tabular}
\end{table*}

\begin{figure*}
\includegraphics[width=7in]{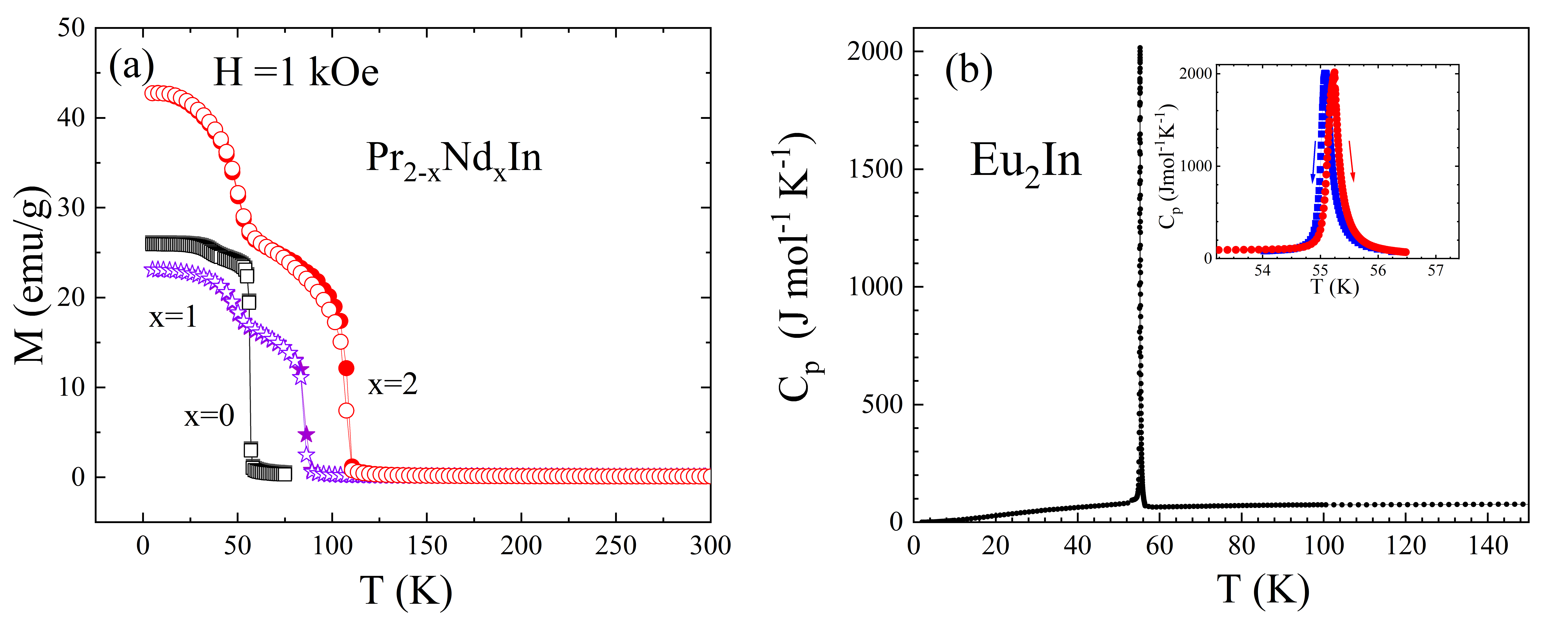} 
\caption {The first-order transition with negligible thermomagnetic hysteresis in Pr$_{2-x}$Nd$_x$In ($x = 0,\,1,\,2$) is evidenced by a sharp transition in the $M$ vs.~$T$ plots shown in (a) \cite{Biswas2020, Biswas2022a, Biswas2022b, Biswas2023}. Open and closed circles represent $M(T)$ data taken during heating and cooling cycles, respectively. (b) The temperature dependence of the specific heat for Eu$_2$In exhibits a pronounced $\delta$-like peak at the transition, indicating a large latent heat. Inset: $C_p(T)$ near the transition during heating and cooling cycles \cite{Guillou2018}.  Figure (b) is reproduced from Ref. \cite{Guillou2018} under Creative Commons Attribution 4.0 International license.}
\label{Fig5}
\end{figure*}

{ \bf \subsubsection*\bf {\noindent ~ B.1. Peculiarities of magnetic phase transitions in R$_2$In with R= light rare-earth atoms (4$f^{<7}$)}}
 
 \paragraph*{\noindent {B.1.a. ~Eu$_2$In:}}\mbox{}\\[0.5em]
First-order magnetic phase transitions (FOMPTs), characterized by their discontinuous nature, are considerably less prevalent in condensed matter than their continuous, second-order counterparts. Nevertheless, they occur across a diverse array of solid materials—including ionic compounds, metals, semimetals, and semiconductors—positioning them as a focal point of contemporary research. These transitions are appealing due to their potential to generate strong non-linear responses to external stimuli, which may lead to remarkable functionalities such as, for example, giant magnetocaloric effect (GMCE) \cite{Roy2013, Trung2010, Wei2015, Biswas2019a, Bez2019}. FOMPTs frequently emerge when magnetic ordering is coupled to structural modifications of the crystal lattice, giving rise to magnetostructural transformations (MST) that are typically accompanied by thermomagnetic hysteresis \cite{Wei2015, Biswas2019a, Bez2019, Fujita2003, Ritter2024}. However, cycling materials through hysteretic MSTs leads to energy dissipation, posing a challenge for practical applications, particularly in energy-related fields like solid-state caloric cooling.

To overcome these limitations, extensive research has focused on engineering materials where magnetic and structural sub-lattices evolve together with minimal hysteresis. Recent advancements suggest that thermomagnetic hysteresis can, in fact, be mitigated, thereby improving the reversibility and efficiency of MST-related functionalities, most notably, the GMCE. Thus far, successful reduction of hysteresis has been largely restricted to transition-metal-based systems \cite{Guillou2018}. In these materials, FOMPTs often involve discontinuous volume changes while retaining crystallographic symmetry, a behavior governed by thermodynamic constraints. One prominent example is the LaFe$_{13-x}$Si$_x$ family of compounds, which exhibit a giant magnetocaloric response with minimal hysteresis \cite{Fujita2003}.  Their transition temperatures are tunable between approximately 200 and 350 K through chemical modification or hydrogenation. The underlying transitions in these and similar compounds are classified as magneto-elastic transformations (METs), closely linked to the unusual characteristics of itinerant-electron metamagnetism (IEM) \cite{Asti1994}. 

In a recent breakthrough, an R$_2$In-type compound Eu$_2$In was found to undergo a FOMPT from a high-temperature paramagnetic (PM) state to a low-temperature ferromagnetic (FM) state at $T_{\mathrm{C}} = 55$~K. This transition is particularly remarkable as it is associated with exceptionally low thermomagnetic hysteresis ($\sim 0.1$~K), earning it the designation of a ``non-hysteretic first-order transition''\cite{Guillou2018}. X-ray absorption near edge spectroscopy (XANES), X-ray magnetic circular dichroism (XMCD), and $^{151}$Eu M\"ossbauer experiments conclusively confirm that Eu exists in a divalent state in Eu$_2$In down to 5~K \cite{Guillou2018, Ryan2019}.

The first-order nature of this transition is clearly evidenced by a pronounced $\delta$-like peak in the specific heat data [see Fig.~5(b)] and large latent heat, $\Delta L \approx 1.4$~kJ/kg. Although the transition is first-order, the crystallographic symmetry remains orthorhombic across the studied temperature range between 5 and 300~K, with relatively minor discontinuous changes in the lattice parameters ($\Delta a/a \approx 0.08\%$, $\Delta b/b \approx 0.04\%$, and $\Delta c/c \approx 0.03\%$) and a slight increase in lattice volume ($\Delta V/V \approx +0.1\%$) during heating at $T_{\mathrm{C}}$ \cite{Guillou2018}. Thus, Eu$_2$In exhibits a unique combination of magnetic and structural characteristics: a first-order magnetoelastic transition with significant latent heat yet minimal thermomagnetic hysteresis---an exceedingly rare combination among rare-earth-based intermetallic compounds. The application of hydrostatic pressure (up to 0.85~GPa) causes a slight positive shift in the transition temperature ($\partial T_{\mathrm{C}}/\partial P \approx 2$~K$\cdot$GPa$^{-1}$). This is because the low-temperature FM phase is the low-volume phase  \cite{Guillou2018}, and the applied pressure therefore stabilizes the FM configuration. Free-energy analysis predicts that the first-order nature of the transition weakens under increasing magnetic field and eventually becomes second-order for fields $H \geq 2.5$~T \cite{Alho2020}.

M\"ossbauer spectroscopy and neutron diffraction studies on Eu$_2$In were also carried out to gain insights into the intriguing magnetism of the compound \cite{Ritter2024}. The M\"ossbauer spectrum of Eu$_2$In at 5~K reveals two components of equal area, indicating that Eu occupies two crystallographic sites of equal multiplicity. However, the distinct hyperfine fields---27~T and 17~T---suggest different magnetic environments for the Eu moments on the two $4c$ sites. Neutron diffraction data collected at 2.5~K confirm that Eu$_2$In exhibits ferromagnetic ordering, with Eu moments on both sites aligned parallel to the $a$-axis. The measured moment values per Eu1 and Eu2 are 6.8~$\mu_{\mathrm{B}}$ and 6.5~$\mu_{\mathrm{B}}$, respectively \cite{Ryan2019}. Guillou \textit{et al.} also reported the influence of Yb substitution on the magnetic properties of Eu$_2$In \cite{Guillou2020}. Substituting Yb at the Eu site in Eu$_2$In leads to a decrease in both the magnetic ordering temperature and the saturation magnetization. Additionally, the transition becomes broader compared to that in pure Eu$_2$In \cite{Guillou2020}.

\begin{figure*}
\includegraphics[width=5.5in]{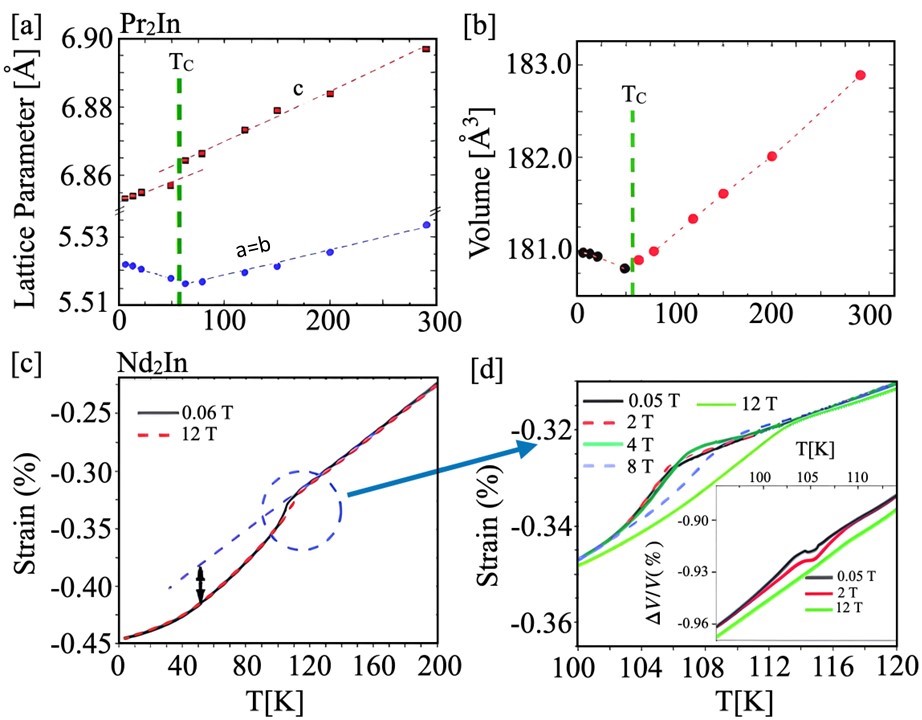} 
\caption {The variations in lattice parameters and unit cell volume across the first-order transition for Pr$_2$In are presented in (a) and (b), respectively \cite{Biswas2022a}. (c, d)  The anomaly in the temperature-dependent longitudinal strain under varying magnetic fields for Nd$_2$In reflects a first-order transition. The inset in (d) illustrates the corresponding volume change \cite{Liu2021}. Figures (c, d) are reproduced from Ref. \cite{Liu2021} under Creative Commons Attribution 4.0 International license. }
\label{Fig6}
\end{figure*}

 \paragraph*{\noindent {B.1.b.~Pr$_2$In and  Nd$_2$In:}}\mbox{}\\[0.5em]
The first-order transition with negligible thermomagnetic hysteresis is also observed in the cases of Pr$_2$In and Nd$_2$In \cite{Biswas2020, Biswas2022a, Biswas2022b, Liu2021, Cui2022}. Unlike Eu$_2$In, both samples adopt the hexagonal Ni$_2$In crystal structure. The rare-earth atoms (Pr and Nd) are believed to be nominally trivalent in both compounds, in contrast to Eu$_2$In where Eu is divalent \cite{Biswas2020, Biswas2022a}.

In Pr$_2$In, a first-order transition from the paramagnetic to the ferromagnetic state occurs at $T_{\rm C} \approx 57$~K, as evidenced by the sharp transition in $M(T)$ [Fig.~\ref{Fig5}(a)] and a $\delta$-like anomaly in the heat capacity data  \cite{Biswas2020}. The sample also exhibits a second-order low-temperature transition at $\sim$35~K [Fig.~\ref{Fig5}(a)], possibly due to spin reorientation \cite{Biswas2020}. The high temperature transition shifts to lower values under the application of hydrostatic pressure ($\partial T_{\rm C}/\partial P \approx -1.9$~K$\cdot$GPa$^{-1}$), exhibiting a rate of change comparable to that observed in Eu$_2$In, although $T_{\rm C}$ decreases in this case. This is because, unlike Eu$_2$In, the FM state is the high-volume phase in the case of Pr$_2$In, i.e., the unit cell volume increases in the FM state (shown and discussed below, see Fig.~\ref{Fig6}) \cite{Biswas2020}.  

The temperature-dependent X-ray diffraction measurements of Pr$_2$In show conventional thermal expansion above $T_{\rm C}$ [Figs.~\ref{Fig6}(a, b)]. However, below $T_{\rm C}$, anomalous behavior emerges: small but distinct discontinuities appear in the unit cell volume and the $c$ lattice parameter, while no discontinuity is detected in the $ab$ basal plane near the transition [Fig.~\ref{Fig6}(a)] \cite{Biswas2022a}. As temperature decreases from room temperature toward $T_{\rm C}$, both the lattice parameter $a$ ($= b$) and the unit cell volume decrease steadily [Fig.~\ref{Fig6}(a)]. Below $T_{\rm C}$, both $a$ ($= b$) and the phase volume increase, whereas the $c$ parameter continues to contract [Fig.~\ref{Fig6}(a)]. The total volume change across the transition is modest---approximately 0.1\%---in agreement with estimates based on the Clausius-Clapeyron relation \cite{Biswas2022a}. Importantly, this symmetry-invariant change in lattice parameters, unit cell volume, and $c/a$ ratio coincides with the magnetic transition [Figs.~\ref{Fig6}(a, b)], providing definitive evidence for a first-order magnetoelastic phase transition in Pr$_2$In \cite{Biswas2022a}. The preservation of crystallographic symmetry and the minimal volume change across the transition explain the negligibly small thermal hysteresis observed in this system.  

Notably, the temperature-dependent magnetization behavior is qualitatively similar for Nd$_2$In [Fig.~\ref{Fig5}(a)]. A sharp transition is observed at $\sim 110$~K with negligibly small thermomagnetic hysteresis followed by a second broad transition at $\sim 50$~K [Fig.~\ref{Fig5}(a)] \cite{Biswas2022b, Liu2021}. The sharp transition at 110~K is believed to be a borderline first-order transition \cite{Biswas2022b, GegenMater25}. The peak in heat capacity data associated with this transition is not $\delta$-like symmetric as typically observed in the case of first-order transitions \cite{Biswas2022b, GegenMater25}. At the same time, Liu \textit{et al.} reported clear changes in lattice strain during the transition [Figs.~\ref{Fig6}(c, d)] \cite{Liu2021}. Their investigations of linear thermal expansion in multiple magnetic fields point out that the phase transition in Nd$_2$In proceeds in two distinct stages, accompanied by negligible volume change. The longitudinal strain initially increases with applied magnetic field, followed by a subsequent decrease [Figs.~\ref{Fig6}(c, d)] \cite{Liu2021}.

The temperature-dependent XRD study shows no change in crystal symmetry for Nd$_2$In during the transition. Moreover, unlike in Pr$_2$In and Eu$_2$In, Nd$_2$In exhibits no discontinuities in either the phase volume or lattice parameters at $T_{\rm C}$. However, the changes in the slopes of the temperature dependence of lattice volume and lattice parameters below $T_{\rm C}$ suggest the presence of weak magnetoelastic coupling. Further analyses of magnetocaloric parameters confirm that the magnetoelastic transition in Nd$_2$In lies on the borderline between first-order and second-order character. It displays certain features typical of first-order transitions, such as sharp magnetic changes \cite{Biswas2022b, Liu2021} and abrupt change in hyperfine field \cite{Forker2005}, but lacks others, such as structural discontinuities \cite{Biswas2022b}.

It is worth noting that for both Pr$_2$In and Nd$_2$In, the experimentally observed saturation magnetization is lower than the theoretically calculated value expected from two symmetry-distinct rare-earth moments with equal magnetic contributions \cite{Biswas2020, Biswas2022b, Liu2021}. This discrepancy likely arises from strong crystal field effects, as indicated by hyperfine interaction measurements \cite{Forker2005}. Additionally, Pr$_2$In exhibits significant magnetic anisotropy, reflected by its large coercivity, whereas Nd$_2$In shows comparatively lower coercivity \cite{Biswas2020, Biswas2022b}. Moreover, the solid solution of PrNdIn samples adopts Ni$_2$In-type hexagonal crystal structure and also shows borderline first-order transition with a magnetic transition temperature of 86~K [Fig.~\ref{Fig5}(a)] \cite{Biswas2023}.

\paragraph*{\noindent B.1.c.~R$_2$In with R = Ce, Sm, and Y:}\mbox{}\\[0.5em]
The compound Sm$_2$In exhibits a ferrimagnetic ground state, setting it apart from other members of the R$_2$In series \cite{Ravot1995}. A magnetic transition from the paramagnetic to ferrimagnetic state occurs at approximately 150~K, as confirmed by muon spin spectroscopy studies \cite{Ravot1995, McAlister1984b}. Notably, McAlister \textit{et al.} \cite{McAlister1984b} reported that the ferrimagnetism in Sm$_2$In is unusual, as its spontaneous magnetization approaches zero upon cooling to the lowest temperatures. Ce$_2$In undergoes a structural phase transition from a hexagonal structure to an as-yet undefined symmetry under hydrostatic pressure of 4.6~GPa, suggesting that further investigation is warranted to determine whether the high-pressure phase adopts an orthorhombic structure. Magnetization studies indicate that Ce$_2$In shows an antiferromagnetic transition at 12~K \cite{Chen1989}. Y$_2$In is paramagnetic and exhibits a structural transition from room temperature Ni$_2$In crystal structure to low temperature Co$_2$Si structure at 250~K \cite{Svanidze2008}.

{ \bf \subsubsection*\bf {\noindent ~ B.2. R$_2$In with R = heavy rare-earth atoms (4$f^{\geqslant7}$)}}

Numerous studies have been reported on the magnetic properties of R$_2$In compounds, where R is a heavy rare-earth element (with 4$f$ electron count $\geqslant$ 7). Magnetism of Gd$_2$In is distinct due to well characterized metamagnetic behavior, while other compounds are less studied. Among the R$_2$In compounds, Gd$_2$In with zero orbital magnetic moment exhibits highest paramagnetic to ferromagnetic transition temperature of $T_{\mathrm{C}} \approx 190$\,K, followed by a transition to an antiferromagnetic state at $T_{\mathrm{N}} \approx 100$\,K \cite{McAlister1984a, GamariSeale1979}.  Pronounced anomalies in the temperature-dependent resistivity near these transitions have also been observed. McAlister \textit{et al.} proposed the spiral antiferromagnetic structure along the $c$-axis for the low-temperature Gd$_2$In phase \cite{McAlister1984a}. Subsequent experimental investigation revealed that the ferromagnetic state of Gd$_2$In below $T_{\mathrm{C}} \approx 190$\,K is actually a helical ferromagnet \cite{Jee1996}. Moreover, the antiferromagnetic state below $T_{\mathrm{N}} \approx 100$\,K emerges only under low magnetic field and undergoes a metamagnetic transition at higher fields \cite{Jee1996}.

Further magnetization studies by Bhattacharya \textit{et al.} reaffirmed these findings, particularly emphasizing the metamagnetic nature of the antiferromagnetic-to-ferromagnetic (AFM–FM) transition below 100\,K \cite{Bhattacharyya2012}. The critical field required for this transition at low temperatures was found to be strongly temperature-dependent. These magnetization results are consistent with hyperfine-field studies conducted by Forker \textit{et al.}, which indicate that the transition near 190\,K is of second-order nature  \cite{Forker2005}. The influence of hydrostatic pressure on both magnetic transitions was examined by Liu \textit{et al.} \cite{Liu1987}. Their findings show that $T_{\mathrm{C}}$ increases slightly under applied pressure, while $T_{\mathrm{N}}$ decreases, suggesting that the antiferromagnetic transition may be completely suppressed below a critical pressure of $\sim$30\,kbar \cite{Liu1987}.

The magnetic behavior of Gd$_2$In is also sensitive to substitution of In by other $p$-block elements. A comprehensive study by Tencé \textit{et al.} demonstrated that doping with Group IVA elements (Sn, Pb) enhances ferromagnetism, as evidenced by a significant increase in $T_{\mathrm{C}}$ (by more than 50\,K) and suppression of the low-temperature antiferromagnetic transition from those obtained for Gd$_2$In \cite{Tence2016}. In contrast, substitution with isoelectronic Group IIIA elements (Al, Ga) had minimal impact on $T_{\mathrm{C}}$, although the antiferromagnetic transition shifted to lower temperatures \cite{Tence2016}. These substitutional effects strongly suggest that 4$f$–5$d$ interactions play a crucial role in magnetic ordering in Gd$_2$In.

In the lanthanide series, Lu$^{3+}$ ion possesses a fully filled 4$f$ shell (4$f^{14}$), and has no unpaired electrons. Consequently, Lu$_2$In is expected to be non-magnetic, i.e., diamagnetic. In the case of Yb$_2$In, analysis based on experimentally obtained lattice parameters suggests that ytterbium is in a divalent state (Yb$^{2+}$; 4$f^{14}$), again leading to a diamagnetic behavior of the compound \cite{Guillou2020}. Further evidence from resistivity, specific heat, magnetic measurements, and temperature-dependent X-ray diffraction studies confirms the absence of magnetic or structural transitions in Yb$_2$In \cite{Guillou2020}.

Starting from the highest atomic number end of the lanthanide series, after Lu and Yb, Tm$_2$In is the first R$_2$In compound to exhibit magnetic behavior. Magnetization measurements reveal a maximum in the temperature-dependent magnetic susceptibility at 17\,K, and the temperature dependent susceptibility obeys the Curie–Weiss law at higher temperatures, with a paramagnetic Weiss temperature of approximately 6\,K \cite{GamariSeale1979, Parviainen1980}.

Magnetic and electrical resistivity measurements, along with Mössbauer spectroscopy and neutron powder diffraction, show that Er$_2$In undergoes a ferromagnetic transition below a critical temperature of $T_{\mathrm{C}} \approx 40$\,K \cite{Ravot1993}. The magnetic moments of Er$^{3+}$ ions at the (2$a$) and (2$d$) crystallographic sites are found to be identical and Mössbauer spectroscopy reports saturation moment per Er atom as $M_{\mathrm{S}} = 8.0 \pm 0.4\,\mu_{\mathrm{B}}$ at 4.2\,K, while neutron diffraction yields $M = 7.7 \pm 0.2\,\mu_{\mathrm{B}}$ at 1.5\,K. These values are slightly lower than the theoretical saturated magnetic moment value for a free Er$^{3+}$ ion ($gJ = 9\,\mu_{\mathrm{B}}$), likely due to the influence of the crystalline electric field (CEF) \cite{Ravot1993}. A second-order ferromagnetic transition at slightly higher $T_{\mathrm{C}} = 46$\,K has been also reported by Zhang \textit{et al.}, underscored by a $\lambda$-type anomaly in the specific heat \cite{Zhang2011}. Above $T_{\mathrm{C}}$, the magnetic susceptibility follows the Curie–Weiss law, with an effective paramagnetic moment close to the theoretical value for a free Er$^{3+}$ ion \cite{Zhang2011}.

For Ho$_2$In, magnetization studies indicate two successive magnetic transitions: a ferromagnetic ordering transition at $T_{\mathrm{C}} \approx 85$\,K, followed by a spin-reorientation transition at $T_{\mathrm{SR}} \approx 32$\,K \cite{Zhang2009b}. These transitions are corroborated by magneto-transport and muon spin rotation ($\mu$SR) measurements \cite{Bhattacharyya2009b, Ravot1997}. Bhattacharya \textit{et al.} also reported the presence of short-range magnetic fluctuations persisting above $T_{\mathrm{C}}$ \cite{Bhattacharyya2009b}. Neutron diffraction studies on Ho$_2$In revealed a collinear ferromagnetic configuration below $\sim$80\,K, with four equivalent magnetic moments on Ho$^{3+}$ ions within the unit cell \cite{Ravot1992}. The orientation of these moments is temperature-dependent: at higher temperatures (40\,K $<$ T $<$ 80\,K) they align along or near the hexagonal $c$-axis; at lower temperatures they gradually rotate toward the basal plane, reaching an inclination of approximately 30$^\circ$ at T = 20\,K. The characteristic temperatures (T $\approx$ 80\,K and 20\,K $<$ T $<$ 40\,K) correspond to the two magnetic transitions observed in the system \cite{Ravot1992}.

\begin{figure*}
\includegraphics[width=5.5in]{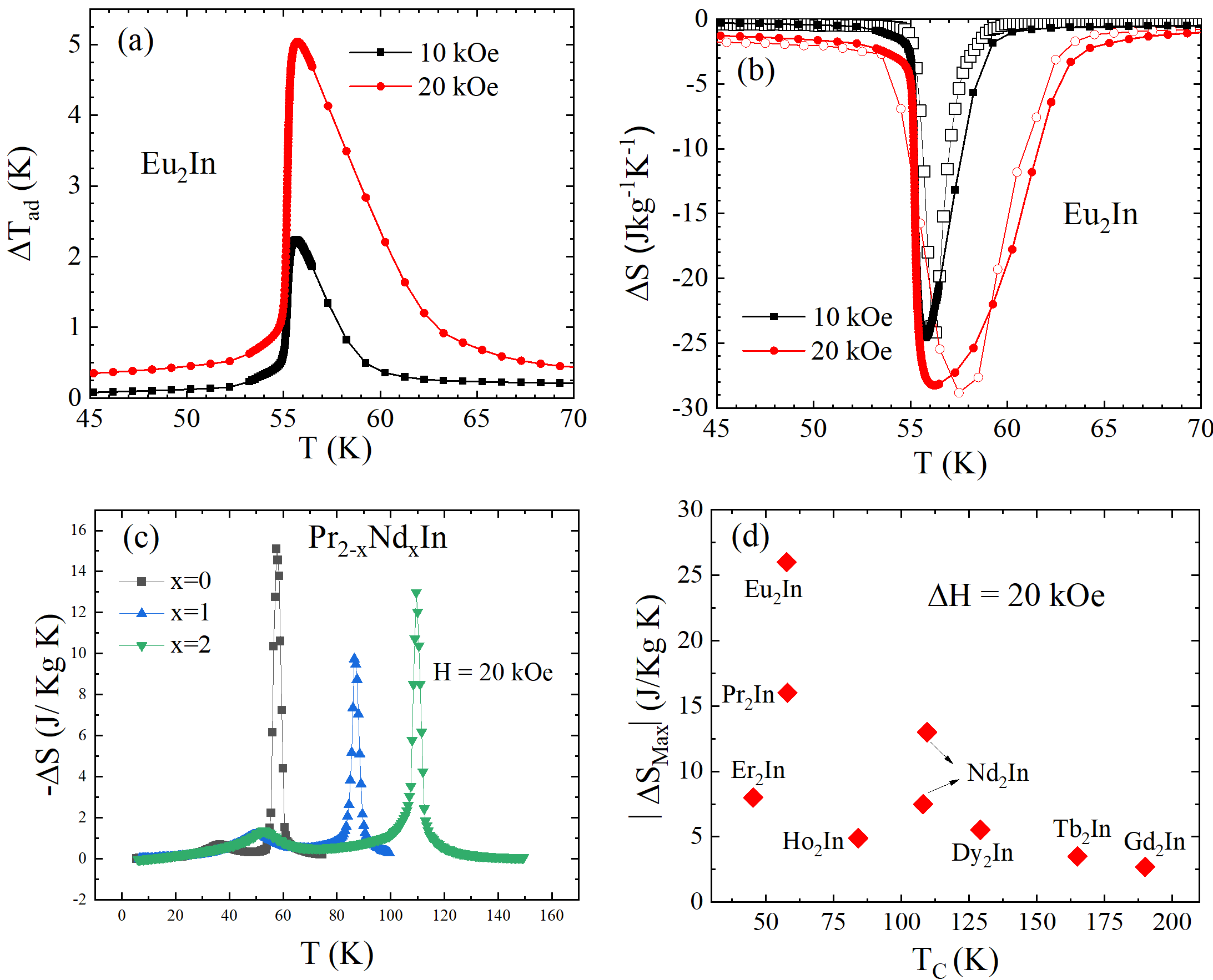} 
\caption {The temperature dependence of $\Delta T_{\rm ad}$ and $\Delta S$ for Eu$_2$In is presented in (a) and (b), respectively. The filled symbols and open symbols curves represent $\Delta S$ calculated from calorimetry and magnetization, respectively. Figures (a, b) are reproduced from Ref. \cite{Guillou2018} under Creative Commons Attribution 4.0 International license. The variation of $\Delta S(T)$ for Pr$_{2-x}$Nd$_x$In at $H = 20$ kOe is shown in (c) \cite{Biswas2020, Biswas2022b, Biswas2023}, while (d) compares the maximum $\Delta S$ near the magnetic transition of different R$_2$In compounds (for $\Delta H=20$ kOe) as a function of their transition temperatures \cite{Guillou2018, Biswas2020, Bhattacharyya2009, Zhang2011, Bhattacharyya2012, Zhang2009c, Bhattacharyya2009b}.}
\label{Fig7}
\end{figure*}

Dy$_2$In shows second-order ferromagnetic transition at $T_{\mathrm{C}} \approx 130$\,K \cite{Zhang2009}, with an additional low-temperature spin-reorientation transition near 50\,K \cite{Bhattacharyya2009}. The second-order nature of the ferromagnetic transition is reflected in a $\lambda$-type anomaly in the specific heat \cite{Bhattacharyya2009}. Dy$_2$In also exhibits significant magnetostriction: near the transition temperature, magnetostriction increases almost linearly with magnetic field, reaching $\sim$100\,ppm at 70\,kOe. Substituting part of the indium with aluminum results in a reduction of the magnetic transition temperature in Dy$_{2}$In$_{1-x}$Al$_x$ \cite{Wang2022}.

Tb$_2$In undergoes two magnetic transitions: a paramagnetic-to-ferromagnetic transition at $T_{\mathrm{C}} \approx 165$\,K, followed by a ferromagnetic-to-antiferromagnetic transition at $T_{\mathrm{N}} \approx 45$\,K \cite{Zhang2009c}. In general, for R$_2$In compounds with heavy rare-earth elements, the ferromagnetic transition temperature ($T_{\mathrm{C}}$) decreases from Gd to Tm with increasing 4$f$ electron count, as noted earlier. Both Gd$_2$In and Tb$_2$In exhibit secondary antiferromagnetic transitions at lower temperatures.

In summary, based on the existing literature, described in this section, it can be concluded that none of the R$_2$In compounds containing heavy rare-earth elements exhibit a first-order transition from the paramagnetic to ferromagnetic state. Instead, their $T_{\mathrm{C}}$ decreases monotonically with increasing atomic number, with Gd$_2$In exhibiting the highest $T_{\mathrm{C}}$ among them. In contrast, several R$_2$In compounds with light rare-earth elements undergo a paramagnetic to ferromagnetic transition, which is first-order in nature. It is interesting to note that two materials that do not have first-order transitions do not order ferromagnetically: Sm$_2$In is ferrimagnetic and Ce$_2$In is paramagnetic. For light rare-earth R$_2$In compounds adopting a hexagonal crystal structure, $T_{\mathrm{C}}$ increases monotonically with increasing atomic number. The most remarkable feature of the light rare-earth R$_2$In compounds, the emergence of first-order transitions accompanied by negligibly small thermomagnetic hysteresis, contrasts with the typical behavior expected for such magnetic transitions.

\subsection{\noindent Magnetocaloric effect of R$_2$In}

While magnetocaloric refrigeration has garnered significant attention for applications near room temperature, its potential advantages in the cryogenic temperature range are equally noteworthy. In particular, the development of energy-efficient methods for liquefying technologically important gases is of growing interest. Notably, R$_2$In compounds show magnetic transitions in the temperature range between 10 and 200~K, which are associated with pronounced magnetocaloric effect.

Among them, Eu$_2$In stands out for exhibiting the largest MCE, primarily not only due to its pronounced first-order magnetic transition occurring near 55~K, but also due to large moment and collinear ferromagnetic structure \cite{Guillou2018}.  The peak adiabatic temperature change ($\Delta T_{\rm ad}$) reaches 2.2~K and 5.0~K for magnetic field changes of 10 and 20~kOe, respectively [Fig. \ref{Fig7}(a)]\cite{Guillou2018}. This large $\Delta T_{\rm ad}$ is particularly remarkable given that, in this temperature regime, the material’s lattice entropy increases steeply, which typically acts to diminish adiabatic temperature change. The emergence of large $\Delta T_{\rm ad}$ places Eu$_2$In among the leading low-temperature giant magnetocaloric materials currently known. The abruptness of the magnetic transition, combined with its high sensitivity to applied magnetic fields ($\partial T_{\rm C}$/$\partial B \approx$ +3.5~K/T \cite{Guillou2018}), ensures that even modest field variations—such as 10~kOe—can convert a significant portion of the latent heat into a magnetocaloric response. Calorimetric analysis reveals exceptionally large magnetic field induced entropy change ($\Delta S$) with the values of $-24.4$ and $-28.2$~J$\cdot$kg$^{-1}\cdot$K$^{-1}$ for the applied magnetic fields of 10 and 20~kOe, respectively. The negligible thermomagnetic hysteresis associated with magnetic transition also renders thermal reversibility of magnetocaloric response, which is highly desirable for magnetic refrigeration \cite{Guillou2018}. The MCE diminishes due to broadening of magnetic transition and reduced magnetization in pseudobinary Eu$_{2-x}$Yb$_x$In and the transition temperature also shifts to lower temperature ranges \cite{Guillou2020}.

Pr$_2$In also exhibits a pronounced magnetocaloric effect (MCE) in the same temperature range as Eu$_2$In, driven by its first-order magnetic transition near 57~K \cite{Biswas2020, Biswas2022a, Cui2022, Liu2024}. For a magnetic field change of 20~kOe, the maximum $\Delta S$ reaches approximately 15~J$\cdot$kg$^{-1}\cdot$K$^{-1}$ \cite{Biswas2020}. Liu \textit{et al.} reported a large $\Delta T_{\rm ad} \sim 2$~K in fields of 2~T and $\sim 4.3$~K in fields of 5~T \cite{Liu2024}. Nd$_2$In demonstrates appreciably large magnetocaloric response near $\sim 110$~K \cite{Biswas2022b, Liu2021}. Based on temperature-dependent magnetization data under varying magnetic fields, the maximum $\Delta S$ is calculated as $-13$~J$\cdot$kg$^{-1}\cdot$K$^{-1}$ for $\Delta H = 20$~kOe at $T_{\rm C} = 110$~K, increasing to $-18$~J$\cdot$kg$^{-1}\cdot$K$^{-1}$ for $\Delta H = 50$~kOe \cite{Biswas2022b}. Although the maximum entropy changes ($\Delta S$) observed for Pr$_2$In and Nd$_2$In are somewhat lower than those reported for Eu$_2$In, they remain comparable to—or even exceed—the values of most known rare-earth based magnetocaloric compounds within the 40–150~K temperature range \cite{Goswami_JPCM_2020, Kumar2024, Chakraborty2024, Kumarprb24, Alho2022, Chakraborty2024}, including other members of the R$_2$In family (see Fig.~\ref{Fig7}(d)). The notably high MCE observed near 110~K in Nd$_2$In is particularly significant, given its proximity to the boiling point of natural gas \cite{Biswas2022b, Liu2021}.  Moreover, the first-order magnetic transitions near $T_{\rm C}$ in these compounds exhibit minimal thermal hysteresis, indicating low hysteresis losses during the refrigeration cycle.

Magnetocaloric behavior has also been investigated in the pseudo-binary series Pr$_{2-x}$Nd$_x$In. Compounds in this series display first-order magnetic transitions accompanied by substantial MCE, with $T_{\rm C}$ spanning from 55~K to 110~K depending on the Nd concentration ($x$) (Fig.~\ref{Fig7}(c)). This temperature range conveniently encompasses the boiling points of various application-relevant gases including nitrogen, oxygen, and natural gas. The intermediate composition PrNdIn demonstrates a maximum entropy change of $-10$~J$\cdot$kg$^{-1}\cdot$K$^{-1}$ for a magnetic field change of 20~kOe \cite{Biswas2023}. The first-order nature of the magnetic transition in Eu$_2$In, Pr$_2$In, and Nd$_2$In is further evidenced by the magnetic field dependence of $\Delta S$. The local exponent $n(H,T) = \frac{d \ln|\Delta S|}{d \ln H}$ exhibits a peak exceeding 2 near $T_{\rm C}$ in all three compounds—a characteristic feature of first-order transitions \cite{Biswas2022a, Biswas2022b, Liu2021, Guillou2020}.

\begin{figure*}
\includegraphics[width=7in]{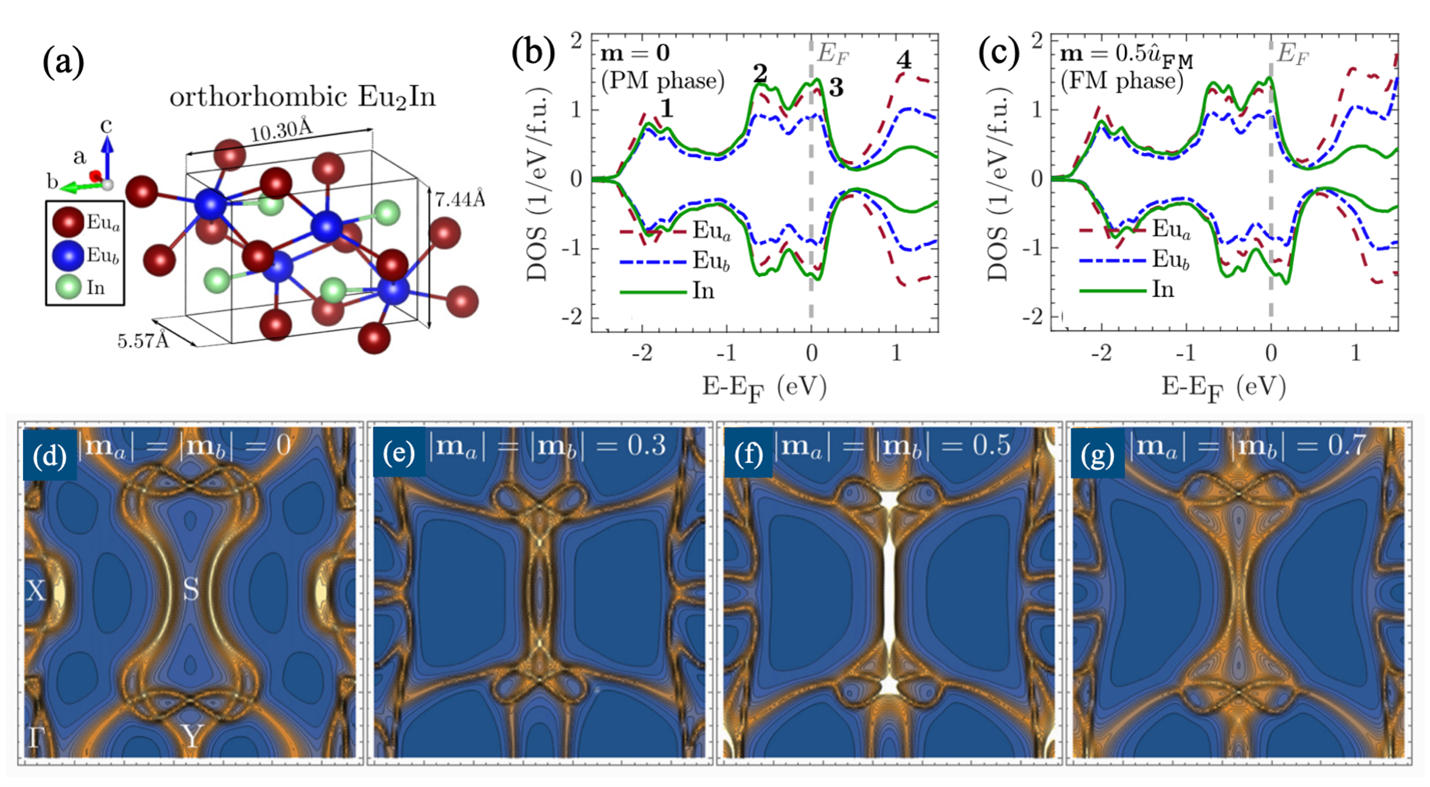} 
\caption { Crystallography and electronic structure of Eu$_2$In.  (a) Orthorhombic unit cell with two inequivalent Eu sublattices (Eu (a), Eu (b)) and In framework.  (b) Sublattice- and orbital-resolved density of states (DOS) in the paramagnetic (PM) state showing symmetric spin character and localized Eu 4$f$ peaks near $E_{\rm F}$.  (c) DOS in the ferromagnetic (FM) state with $m = 0.5\,\mu_{\rm F}$, where exchange splitting redistributes spectral weight and induces spin polarization via Eu–In hybridization.  (d–g) Bloch spectral functions at $E_{\rm F}$ for majority-spin electrons, illustrating the transition from incoherent PM spectra to sharp FM bands with reconstructed Fermi surfaces. Figure is reproduced from Ref. \cite{MendiveTapia2020} under Creative Commons Attribution 4.0 International license. }
\label{Fig8}
\end{figure*}

A moderate magnetocaloric effect (MCE) is observed in Gd$_2$In near its ferromagnetic transition temperature around 190~K \cite{Bhattacharyya2012}. Additionally, an inverse MCE appears around 100~K under low magnetic fields, associated with the transition from the ferromagnetic to the antiferromagnetic state. At higher magnetic fields, the entropy change at low temperature becomes negative, consistent with conventional MCE behavior \cite{Bhattacharyya2012}. Other R$_2$In compounds containing heavy rare-earth elements also exhibit magnetic transitions accompanied by modest MCE values—comparable to those of Gd$_2$In but lower than the MCE observed in light rare-earth analogs that undergo first-order magnetic transitions (see Fig.~\ref{Fig7}(d)) \cite{Bhattacharyya2009, Zhang2009, Zhang2011, Bhattacharyya2009b, Zhang2009b, Zhang2009c}.

A few studies have explored the substitution of other p-block elements at the In site in R$_2$In compounds. Tencé \textit{et al.}  investigated the magnetocaloric effect (MCE) in Gd$_2$In$_{0.8}$X$_{0.2}$ (X = Pb, Al) \cite{Tence2016}. Compared to Gd$_2$In, the Pb-doped sample exhibited a reduced magnetic entropy change, despite an increase in the Curie temperature. In contrast, Al doping had a negligible effect on both MCE and $T_{\rm C}$ \cite{Tence2016}. Additionally, a combined magnetocaloric and magnetostriction study was conducted on Dy$_2$In$_{1-x}$Al$_x$ ($x = 0, 0.2, 0.3, 0.4$) alloys, with Dy$_2$In$_{0.7}$Al$_{0.3}$ showing the highest MCE and magnetostriction \cite{Fornasini1990}. 

To summarize, from a magnetocaloric perspective, light rare-earth-based R$_2$In compounds exhibit a more pronounced MCE due to their first-order magnetic transitions, in contrast to the heavy rare-earth variants. This is unusual because typically MCE of heavy rare-earth compounds is expected to be higher due to their large magnetic moments.  However, it appears that even modest magnetic field is sufficient to convert a substantial portion of the latent heat associated with the first-order transition into a magnetocaloric response in case of light rare-earth based R$_2$In \cite{Guillou2018}.  Notably, the magnetic transitions in R$_2$In compounds—regardless of being first- or second-order—are typically associated with negligible thermal hysteresis, a highly desirable characteristic for potential magnetic refrigerant materials.

\subsection{\noindent ~Theoretical studies }

DFT and related modeling approaches have been widely employed to uncover how $f$-electron valence fluctuations, exchange interactions, and magnetoelastic coupling govern phase stability and magnetic transitions in R$_2$In compounds. These theoretical efforts highlight that subtle electronic and structural factors can produce unusual transition characteristics, such as sharp but reversible first-order behavior. Eu$_2$In stands out as the prototypical example where these mechanisms manifest most clearly, displaying a non-hysteretic first-order magnetic transition with large latent heat and a giant magnetocaloric effect \cite{Guillou2018}.

DFT calculations were employed to understand the electronic structure of Eu$_2$In \cite{Guillou2018, MendiveTapia2020}. Guillou \textit{et al.} in their work point out that the Eu 6$s$, 6$p$, and 5$d$ states are strongly hybridized with the more populated In 5$p$ states, such that the Eu $s$, $p$, $d$ states lie entirely within the envelope of the In-$p$ states \cite{Guillou2018}. Owing to this strong hybridization, the Eu-5$d$ states in Eu$_2$In are partially occupied ($\sim 0.8$ eV per Eu) rather than empty, as would be expected for Eu$^{2+}$ in the $4f^7 5d^0$ configuration. The Eu-4$f$ states are centered just below $-2$ eV and also hybridize with the In $p$ states and, consequently, with the Eu $s$, $p$, $d$ states. These Eu-4$f$ states lie deeper in energy than in most Eu–TM–X ternaries (where TM is a transition metal and X is a $p$-block element), consistent with the absence of mixed valence or valence transitions in Eu$_2$In \cite{Guillou2018}. In the FM state, the Fermi level ($E_{\rm F}$) lies at the edge of a pseudo-gap in the majority channel, while in the minority channel it corresponds to high Eu-5$d$ and In-5$p$ density of states (DOS). In the approximated PM state, $E_{\rm F}$ falls inside a shallow valley flanked by high Eu-5$d$ and In-5$p$ DOS. High DOS near $E_{\rm F}$ is often associated with electronic and magnetic instabilities triggered by external parameters such as temperature, magnetic field, or pressure. The electronic structures capable of large DOS changes at or near $E_{\rm F}$ in Eu$_2$In are attributed to the magnetoelastic first-order magnetic transitions in this compound \cite{Guillou2018}.

In a subsequent DFT study, Mendive-Tapia \textit{et al.} demonstrated that Eu$_2$In exhibits a unique positive feedback between the magnetism of itinerant valence electrons and the ferromagnetic ordering of localized $f$-moments, manifested as a topological change in the Fermi surface that drives its first-order transition \cite{MendiveTapia2020}. As mentioned above, Eu$_2$In crystallizes in an orthorhombic structure [Table~\ref{tab:structure}, Fig.~\ref{Fig8}(a)] with two inequivalent Eu sublattices, Eu(a) and Eu(b), embedded within an In framework. This structural inequivalence enables sublattice-dependent magnetic behavior that becomes crucial in the transition from the paramagnetic (PM) to the ferromagnetic (FM) state [Fig.~\ref{Fig8}(b,c)]. In the PM phase, described by the disordered local moment (DLM) model, the density of states (DOS) shows strong localized Eu 4$f$ features near the Fermi level, while In contributes weakly through its $s$–$p$ orbitals. The spin-symmetric DOS reflects statistical averaging of disordered spins, with hybridization evident between Eu $d/f$ and In $s/p$ states \cite{MendiveTapia2020}.

Upon entering the FM state, exchange splitting separates the Eu sublattices into majority and minority spin channels, redistributing spectral weight near the Fermi level and inducing spin asymmetry through Eu–In hybridization. Bloch spectral functions [Fig.~\ref{Fig8}(d–g)] capture this evolution, revealing a transformation from a broad, incoherent PM spectrum to a sharp, spin-polarized metallic state characterized by band sharpening, symmetry breaking, and Fermi surface reconstruction. Mendive-Tapia \textit{et al.} attributed this behavior to a feedback loop between itinerant carriers and localized Eu $f$-moments—enabled by Eu’s divalency and crystallographic inequivalence—that stabilizes ferromagnetism and drives the first-order transition, a mechanism absent in trivalent Gd$_2$In \cite{MendiveTapia2020}.

\begin{figure*}
\includegraphics[width=7in]{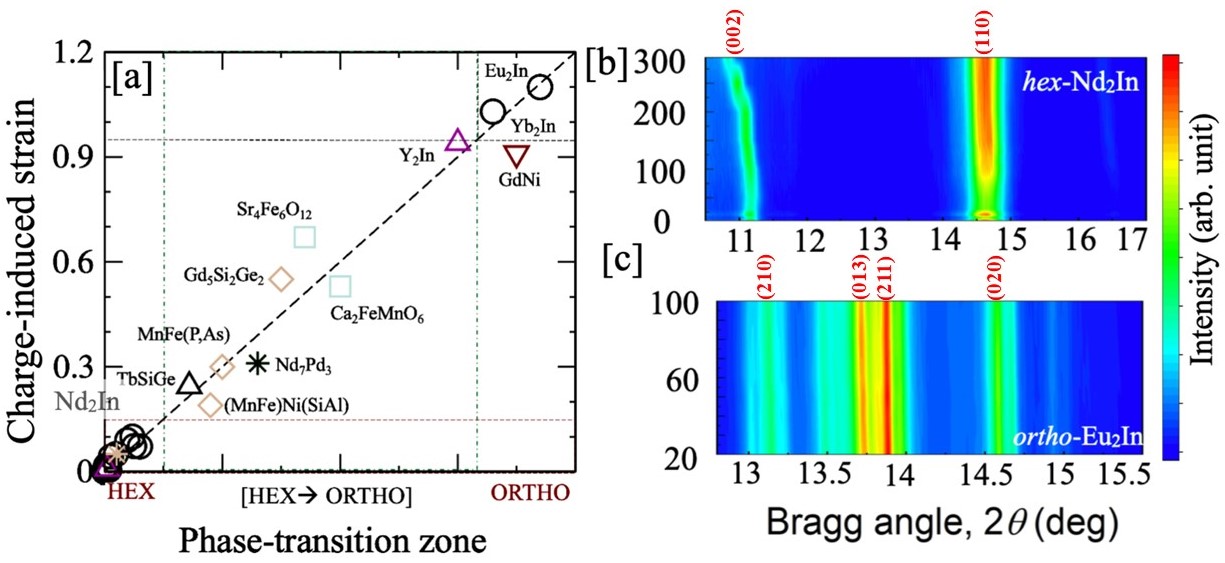} 
\caption {(a) Charge-induced strain as a descriptor for phase prediction, identifying stability zones for hexagonal, transitional (hex$\to$ortho), and orthorhombic phases. (b, c) Experimental validation by temperature-dependent X-ray diffraction (XRD) showing stable hexagonal Nd$_2$In at low strain, while orthorhombic Eu$_2$In at high strain. Figure is reprinted  with permission from \cite{Singh2025}, Copyright (2025) by the American Physical Society.}
\label{Fig9}
\end{figure*}

Alho \textit{et al.} developed a theoretical model to describe the magnetic and magnetocaloric properties of Eu$_2$In \cite{Alho2020}. Their model incorporates both exchange and magnetoelastic interactions within the mean-field approximation, while the coupling between magnetic and lattice entropies is treated using the Grüneisen approach, which relates the Debye temperature to lattice deformation. The study revealed the existence of a critical magnetic field above which the compound’s first-order transition becomes second order. By examining the variation of magnetic free energy with magnetization and temperature, the authors explained the nonhysteretic first-order transition observed in Eu$_2$In. The model not only reproduces the experimentally observed GMCE in this R$_2$In compound but also provides a key insight: despite the relatively small phase volume change, lattice distortion accounts for nearly one-third of the total magnetic-field-induced entropy change ($\Delta S$) in Eu$_2$In. In contrast, the maximum adiabatic temperature change ($\Delta T_{\rm ad}$) remains unaffected by the associated phase volume change \cite{Alho2020}.

After the discovery of a first-order transition in Eu$_2$In, where Eu is divalent, similar first-order transitions were later observed in Pr$_2$In and Nd$_2$In—two R$_2$In-type compounds in which the rare-earth elements are trivalent \cite{Biswas2020, Biswas2022b}. Unlike Eu$_2$In, however, these compounds crystallize in a hexagonal structure. To gain insight into their behavior, DFT-based electronic structure calculations were carried out \cite{Biswas2020, Biswas2022a, Biswas2022b}. The results indicate that, in the majority DOS of Pr$_2$In, the Pr 4$f$ states strongly hybridize with Pr 5$d$ and In 5$p$ states near $E_{\rm F}$, with a pronounced $d$-character \cite {Biswas2022b}. This suggests that the 5$d$ electrons play a significant role in mediating magnetic interactions between local 4$f$ moments. In contrast, for Nd$_2$In, the hybridization between Nd 4$f$, Nd 5$d$, and In 5$p$ states occurs slightly above $E_{\rm F}$, signaling a weakening of the 5$d$ and 5$d$–5$p$ contributions near $E_{\rm F}$. The strong lanthanide 5$d$–indium 5$p$ hybridization with predominantly $d$-character near $E_{\rm F}$ for Pr$_2$In is assumed to drive the first-order transition in the compound in analogy to Eu$_2$In, where linkage of similar electronic feature with first-order transition was established by Mendive-Tapia \textit{et al.} \cite{MendiveTapia2020}. It was also hypothesized that the lack of $d$ contribution near $E_{\rm F}$ due to the occurrence of hybridized states slightly above $E_{\rm F}$ for Nd$_2$In makes the compound exhibit a weaker first-order character of magnetic transition, making it borderline between first-order and second-order. The lack of hybridized states of $d$-character in Nd$_2$In was also reflected via different Fermi surface topology of the compound compared to Pr$_2$In (Fig.~\ref{Fig8}) \cite{Biswas2022b}.

First-principles electronic structure studies of R$_2$In compounds consistently point to the crucial role of orbital hybridization in shaping their magnetic behavior. In particular, the interaction between localized 4$f$ states of the rare-earth elements and the more delocalized 5$d$ states, together with the In 5$p$ conduction states, produces a pronounced $d$-character near the Fermi level. This hybridization not only mediates the exchange interactions responsible for long-range magnetic ordering but also tunes the balance between first-order and second-order magnetic transitions across the series. In compounds such as Eu$_2$In, where $f$–$d$–$p$ hybridization is particularly strong and sensitive to subtle changes in electronic occupancy and lattice volume, the result is a sharp yet reversible first-order transition that gives rise to a giant magnetocaloric effect. By contrast, in heavier lanthanide members, weaker or less favorable hybridization channels shift the balance toward conventional second-order transitions with more moderate caloric responses. Thus, the degree and nature of $f$–$d$–$p$ hybridization at the Fermi level provide a unifying framework for understanding the diversity of magnetic ground states and transition characteristics observed across the R$_2$In family.

Among promising systems, R$_2$In offer a compelling platform as their complex crystallographic tendencies, strong electronic correlations, and sensitivity to external stimuli make them rich hosts of structural and magnetic phase transitions. Recently, Singh \textit{et al.} \cite{Singh2025} built a physics-informed approach that links charge-induced atomic strain with the stability of R$_2$In phases, which provides a path toward uncovering predictive descriptors for phase transitions and guiding the discovery of next-generation functional materials.

Figure~\ref{Fig9}(a) shows charge-induced strain plotted against the phase-transition zone, revealing three distinct regions. Low strain values ($<0.15$) correspond to stable hexagonal phases, such as Nd$_2$In. Intermediate strain values (0.15–1.0) indicate systems prone to hexagonal-to-orthorhombic transitions, exemplified by (Mn$_{0.5}$Fe$_{0.5}$)Ni(Si$_{1-x}$Al$_x$) \cite{Biswas2019a}. High strain values ($>1.0$) correspond to stable orthorhombic phases, including Eu$_2$In, Yb$_2$In, and Y$_2$In. A dashed guideline highlights the progression from hexagonal to orthorhombic stabilization as strain increases, emphasizing its role as a predictive descriptor of structural phase transitions in R$_2$In and other magnetic compounds.  Figure~\ref{Fig9}(b,c) provide experimental validation through temperature-dependent XRD intensity maps. Notably, Nd$_2$In retains hexagonal reflections across temperature, consistent with its low strain and hexagonal stability, while Eu$_2$In exhibits orthorhombic reflections, confirming the prediction of a large strain and an orthorhombic ground state \cite{Singh2025}.  

Although these strain-based descriptors apply primarily to charge-driven displacive transformations in rare-earth intermetallics (such as R$_2$In compounds), they do not necessarily extend to purely reconstructive transitions or entropy-driven, order–disorder changes. In systems lacking appreciable charge-induced distortion, phase behavior must instead be rationalized through alternative mechanisms — such as configurational entropy effects or bond-rearrangement pathways. However, this recent study \cite{Singh2025} opens up a promising area for future studies of complex phase transitions in intermetallic solids.

\section{\noindent ~Summary, Outlook, and Future directions}

In this review, we summarize recent experimental and theoretical advances in the study of binary rare-earth R$_2$In intermetallics. These compounds display unusual magnetic behaviors and phase transitions that are highly sensitive to external stimuli, making them both fundamentally intriguing and promising for technological applications. While first-order magnetic transitions are generally accompanied by significant thermomagnetic hysteresis, several R$_2$In compounds, including Eu$_2$In, Pr$_2$In, and Nd$_2$In, exhibit first-order transitions with negligible hysteresis, an uncommon and striking phenomenon that challenges conventional understanding of magneto-structural coupling. Structurally, most R$_2$In compounds stabilize in the hexagonal Ni$_2$In-type lattice, including the non-lanthanide Y$_2$In. Exceptions are Eu$_2$In and Yb$_2$In, which adopt the orthorhombic Co$_2$Si-type structure. Notably, no R$_2$In compound has been observed to undergo a symmetry-changing structural transition, except Y$_2$In, which transforms from hexagonal to orthorhombic near 250~K. From a theoretical perspective, charge-induced lattice distortions and electronic structure considerations, such as Fermi-surface nesting and sublattice-specific hybridization, provide predictive descriptors for these structural tendencies. In particular, strain-based metrics derived from DFT calculations have successfully rationalized hexagonal-to-orthorhombic stability trends, offering a quantitative framework for predicting phase transitions in this family.  

Magnetically, Gd$_2$In exhibits the highest $T_{\mathrm{C}}$ across the series. Among hexagonal compounds, $T_{\mathrm{C}}$ decreases with increasing atomic number for heavy rare-earth elements, while the opposite trend occurs for lighter members. The choice of rare-earth element R strongly determines the magnetic ground state, but In also mediates inter-sublattice interactions in some compounds. Crystal-field effects, coupled with electronic correlations, further shape the overall magnetic behavior. Many R$_2$In compounds display pronounced MCE in the cryogenic range, encompassing the boiling points of several technologically relevant gases. Eu$_2$In stands out for its exceptionally large MCE, placing it among giant cryogenic magnetocaloric materials. Significant magnetic entropy changes are also reported in Pr$_2$In and Nd$_2$In, while heavy lanthanide-based members (Er$_2$In, Ho$_2$In, Tb$_2$In, Dy$_2$In, and Gd$_2$In) exhibit moderate MCE. Despite extensive research, several critical questions remain. Most notable among them is the mechanism underlying non-hysteretic first-order transitions in light rare-earth compounds, particularly the origin of the large latent heat in Eu$_2$In despite a minimal lattice volume change (0.1\%). Advanced theoretical tools, including combined DFT and statistical modeling, suggest that subtle charge redistribution, local lattice distortions, and spin-lattice coupling may collectively stabilize these transitions without significant hysteresis. Understanding these mechanisms could establish design principles to minimize thermomagnetic hysteresis, a key limitation in practical giant magnetocaloric materials such as Gd$_5$Si$_2$Ge$_2$, Mn$_{0.5}$Fe$_{0.5}$NiSi$_{1-x}$Al$_x$, FeRh, and Ni$_{50-x}$Co$_x$Mn$_{35}$Ti$_{15}$ \cite{Pecharsky1997, Biswas2019a, Bez2019, Biswas_Jalcom_22}.  

Future directions include exploration of pseudobinary systems R$_{2-x}$R$'_x$In, which remain largely understudied. These materials offer a unique platform to investigate the interplay between composition, electronic structure, lattice stability, and magnetism, as well as the effects of multiple rare-earth ions on spin interactions. Mixing 4$f$ elements expands the chemical and magnetic design space, potentially enabling emergent behaviors and novel phase transitions that are highly sensitive to external stimuli. Additionally, integrating theoretical descriptors such as strain, charge redistribution, and electronic structure into predictive frameworks could accelerate discovery of R$_2$In-based materials with tunable phase behavior and optimized magnetocaloric performance. Finally, while R$_2$In compounds show significant potential for cryogenic MCE, chemical instability under ambient conditions---particularly in light rare-earth members---remains a major barrier to practical applications. Future research must focus on enhancing material stability, leveraging both theoretical guidance and experimental synthesis strategies, to enable breakthroughs in efficient, high-performance cryogenic magnetocaloric materials.

\section{Data availability statement}

Data availability statement All data that support the findings of this study are included within the article (and any supplementary files).

\section{Acknowledgments}

The work at Ames National Laboratory was supported by the U.S. Department of Energy, Office of Science, Basic Energy Sciences, Materials Science and Engineering Division. The research is performed at the Ames National Laboratory, which is operated for the U.S. DOE by Iowa State University under contract No. DE-AC02-07CH11358.

\end{document}